# Three-dimensional atom-by-atom measurement of interface diffusion dynamics

Doojin Park[1,†], Hyesung Jo[1,†], Yongmin Kwon[2,3], Seokjo Hong[1], Jaewhan Oh[1], Youngjoo Whang[2,3], Eunjik Lee[3,4,5], Gu-Gon Park[3,4,5], Sang Woo Han[2,*] and Yongsoo Yang[1,6,**]

[1] *Department of Physics, Korea Advanced Institute of Science and Technology (KAIST), Daejeon 34141, Republic of Korea*
[2] *Department of Chemistry, Korea Advanced Institute of Science and Technology (KAIST), Daejeon 34141, Republic of Korea*
[3] *Hydrogen Fuel Cell Laboratory, Korea Institute of Energy Research (KIER), Daejeon 34129, Republic of Korea*
[4] *Department of Energy Engineering, University of Science and Technology (UST), Daejeon 34113, Republic of Korea*
[5] *Graduate School of Energy Science and Technology (GEST), Chungnam National University, Daejeon 34134, Republic of Korea*
[6] *Graduate School of Semiconductor Technology, School of Electrical Engineering, Korea Advanced Institute of Science and Technology (KAIST), Daejeon 34141, Republic of Korea*

[†] These authors contributed equally to this work.
Corresponding author email: *sangwoohan@kaist.ac.kr, **yongsoo.yang@kaist.ac.kr

**Diffusion at interfaces governs how materials and devices form, evolve and function. It is usually measured through smooth concentration profiles and bulk transport coefficients, which average over the discrete atomic rearrangements. At nanoscale interfaces only a few atomic layers wide, however, such coarse-graining obscures the local three-dimensional (3D) atomic transport, and scalar diffusion coefficients cannot fully describe species-, layer- and facet-resolved redistribution. The essential observable is the discrete 3D redistribution of atoms themselves, beyond an averaged profile. Here we combine micro-electro-mechanical systems (MEMS) pulse–quench heating with atomic electron tomography (AET) to reconstruct five 3D atomic configurations of the same coherent Pd@Pt core–shell nanoparticle, each corresponding to a well-defined thermal state. Species-conserving one-to-one assignment links neighbouring configurations, and labelled reference simulations correct the displacement bias caused by indistinguishable same-species atoms. The corrected Pd diffusivities follow Arrhenius behaviour, giving an apparent activation energy of 1.10 ± 0.09 eV, consistent with migration-limited vacancy-mediated transport. Atomic-layer transition statistics further show that increasing 3D mobility can coexist with declining net chemical transfer and reveal facet-dependent pathways hidden by the scalar diffusivity. Together, these measurements establish a 3D atom-by-atom framework that connects identity-corrected mobility to species-, layer- and facet-resolved transport at buried nanointerfaces.**

## Main

For more than a century, diffusion has been described through a compromise between microscopic motion and macroscopic fields. Fick's laws describe fluxes driven by concentration gradients[1], while irreversible thermodynamics extends this picture to chemical-potential gradients[2,3]. Einstein linked diffusion to atomic mean-squared displacement[4,5]. In multicomponent solids, the Kirkendall–Darken framework links tracer motion, chemical thermodynamics, vacancy-mediated lattice-frame motion and interdiffusion coefficients measured from concentration profiles[6–9]. These approaches are most directly applicable in bulk diffusion couples, where many atomic jumps can be averaged over volumes large enough to define smooth fields, reference frames and transport coefficients.

Nanoscale interfaces challenge this compromise. In semiconductor contacts, tunnel junctions, thin-film heterostructures and core–shell nanoparticles, the functional region can be only a few atomic layers wide. On this length scale, a diffusion coefficient inferred from a continuum concentration profile cannot simply be assumed to describe local interfacial transport[10–14]. A quantitative description at this scale therefore requires discrete 3D atom-by-atom transport metrology rather than continuum profile fitting alone.

Existing methods have provided only partial access to atomistic interfacial transport. Radiotracer and interphase-boundary methods quantify diffusion sensitively but infer transport from isotope or concentration profiles[5,15]. Field-ion and atom-tracking scanning tunnelling microscopies resolve individual surface jumps, but not transport across buried interfaces in 3D[16,17]. In-situ electron microscopy can image chemical redistribution during heating, but most observations are projected, beam-sensitive or analysed through morphology and concentration contrast rather than one-to-one atomic transport[13,18–29].

AET offers a route past the projection problem by recovering 3D atomic coordinates and chemical species in individual nanocrystals[30–35]. Anneal-and-image tomography with slow thermal ramps can capture atomic configurations after heating[32,35], but each configuration integrates diffusion over the full temperature history, including heating and cooling, rather than a well-defined isothermal exposure. Moreover, same-species atoms are indistinguishable between reconstructions, so successive snapshots cannot be converted directly into atomic displacements without a physically constrained correspondence and correction for identity loss. Quantitative measurement therefore requires both well-defined thermal states and a calibrated one-to-one correspondence between them.

Here we combine MEMS pulse–quench heating with AET to establish such a measurement. A coherent face-centred cubic (fcc) Pd@Pt core–shell nanoparticle is driven through a sequence of sharply defined thermal states and reconstructed in 3D after each quench. The preserved lattice enables species-conserving one-to-one assignment between neighbouring thermal states[36], while labelled reference simulations calibrate the displacement loss caused by same-species indistinguishability. The resulting measurements combine identity-corrected 3D diffusivities with direct species-, layer- and facet-resolved transition statistics, revealing how total atomic mobility and composition-changing transport can evolve differently at a buried interface.

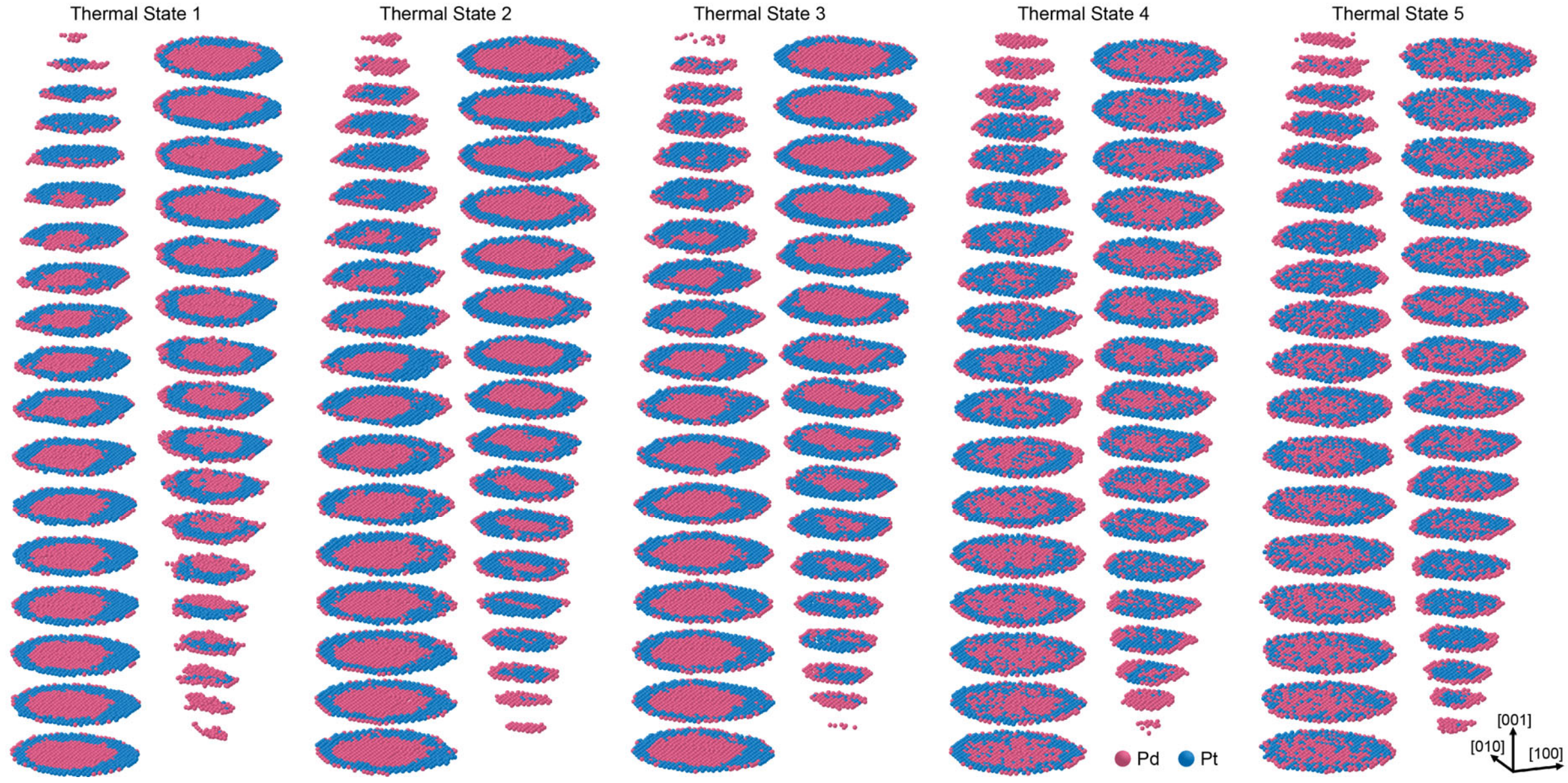

**Figure 1 | Temperature-dependent 3D atomic configurations of a single Pd@Pt nanoparticle.** AET reconstructions of the same Pd@Pt core–shell nanoparticle after sequential MEMS pulse–quench annealing at 300, 400, 500, 600 and 650 °C, corresponding to thermal states 1 to 5, respectively. Atomic models are shown as slices along the [001] direction, with Pd and Pt atoms rendered in magenta and blue, respectively.

## Pulse–quench AET captures atomic configurations during interfacial diffusion

Preliminary pulse–quench annular dark-field scanning transmission electron microscopy (ADF-STEM) experiments identified sequential 900 s annealing steps at 300, 400, 500, 600 and 650 °C as conditions that activate interfacial intermixing while preserving the overall nanoparticle structure (Methods and Supplementary Fig. 1). We selected an isolated Pd@Pt nanoparticle and followed the same particle through these five thermal states. At each step, the particle was rapidly heated within the microscope column under high vacuum, held at the target temperature for 900 s, quenched to room temperature within a few seconds and then measured by AET. The rapid thermal response of the MEMS chip limited diffusion outside the isothermal dwell, so that each reconstruction represents a sharply defined quenched state rather than a thermally blurred annealing history[37] (Methods and Supplementary Fig. 2). AET reconstructs the 3D coordinates and chemical species of individual atoms from a tilt series of ADF-STEM projections[31,32,38,39]. Because the selected particle was isolated and showed no evidence of meaningful mass exchange or ripening, the Pd and Pt atom populations were constrained during the AET chemical classification (Methods).

Figure 1 shows representative [001] slices through the resulting 3D atomic models. Each model contains 6,414 Pd atoms and 6,381 Pt atoms. The particle retains a coherent fcc lattice throughout the thermal sequence, whereas the initially distinct Pd core and Pt shell become progressively intermixed. This common lattice provides a fixed 3D framework for comparing atom-by-atom redistribution between successive thermal states.

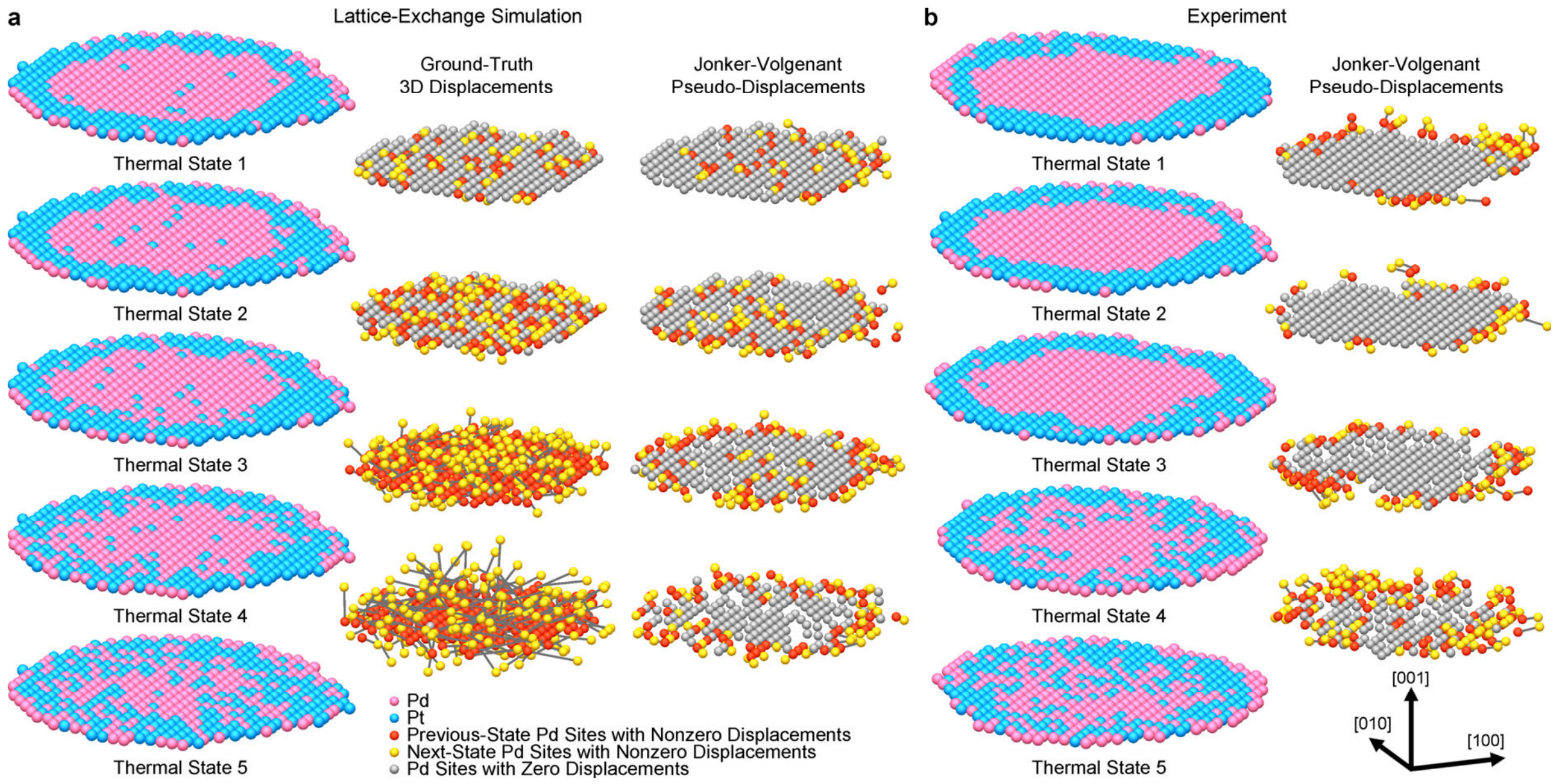

**Figure 2 | Simulation calibration and experimental pseudo-trajectories. a**, Composition-conserving lattice-exchange configurations matched to thermal states 1–5. For each neighbouring thermal state pair, retained simulation labels provide the ground-truth 3D Pd displacements, whereas the Jonker–Volgenant pseudo-trajectories show a systematically compressed displacement distributions. **b**, Experimental AET configurations and the corresponding Jonker–Volgenant Pd pseudo-trajectories for neighbouring thermal state pairs. Magenta and blue spheres denote Pd and Pt, respectively. In the displacement maps, red and yellow denote the same displaced Pd atom, or Jonker–Volgenant-matched Pd atom pair, in the previous and next thermal states, respectively; grey denotes Pd atoms with no displacement between the two states. Lines connect the corresponding displaced endpoints.

## Simulation-calibrated atom matching

The AET reconstructions provide experimental 3D atomic coordinates and chemical species, but without atom identities across thermal states. We therefore conducted composition-conserving nearest-neighbour exchange simulations which generated sequences of labelled configurations (Methods). The simulated states corresponding to experimental thermal states were selected by matching their layer-resolved Pd distributions to the measured distributions (Methods). Although the exact sequence of atomic movements between the measured snapshots cannot be determined or simulated exactly (Pd–Pt intermixing may involve temperature-, morphology-, surface- and/or vacancy-dependent processes[18,40]), composition-conserving exchanges can still be utilized for sampling configuration snapshots during the diffusion[41]. Here, the exchange simulations are used only to quantify assignment bias by comparing the same simulated snapshot pairs with and without atom-identity labels.

Because identities are retained in the simulation, the ground-truth 3D displacements between neighbouring thermal states are known. After erasing the labels, we paired same-species atoms by minimizing the squared Euclidean distance under a global one-to-one constraint using the Jonker–Volgenant algorithm[36]. The resulting pseudo-trajectories are minimum-total-squared-distance endpoint correspondences between two unlabelled snapshots, not the actual paths taken by the atoms. Figure 2a compares the pseudo-trajectories with the labelled ground-truth displacements and shows systematic displacement compression caused by identity loss, as further supported by the corresponding displacement histograms (Supplementary Fig. 3).

For each species and interval, we therefore defined a multiplicative correction factor $C = \frac{D^{sim}_{labelled}}{D^{sim}_{pseudo}}$ from Einstein-type diffusivities calculated for the same simulated snapshot pair using the labelled ground-truth displacements ($D^{sim}_{labelled}$) and the pseudo-displacements ($D^{sim}_{pseudo}$), respectively. An ensemble of correction factor sets was obtained by independent simulations with chemical-label perturbations (Methods), and the ensemble provides a species- and interval-specific calibration from experimentally measurable pseudo-diffusivities to identity-corrected 3D diffusivities. Correction factors obtained from a separate simulation using an ideal spherical core–shell geometry were consistent with those from the experiment-based geometry, and both simulations showed near-Arrhenius diffusivity behaviour, supporting the robustness of the simulation-based calibration (Supplementary Fig. 3g–i).

## Identity-corrected 3D interfacial diffusivity

Applying the same-species Jonker–Volgenant assignment to consecutive experimental thermal states yielded 3D pseudo-displacements for Pd and Pt over each thermal-state interval (Fig. 2b and Supplementary Fig. 3). Surface atoms were excluded from the Jonker–Volgenant assignment, and assignments within large connected displacement clusters associated with particle reshaping were removed from the analysis (Methods). Einstein-type pseudo-diffusivities were calculated from the resulting 3D mean-squared pseudo-displacements. For each species, each complete correction factor set was applied across all four intervals to obtain $D^{exp}_{corr} = C\ D^{exp}_{pseudo}$, and the resulting four diffusivities were fitted to the Arrhenius relation $D = D_0 \exp\left(-\frac{E_a}{k_B T}\right)$. Repeating this procedure for each correction factor set yielded paired $E_a$ and ln $D_0$ values, whose ensemble means and standard deviations define the reported estimates and calibration-derived uncertainties (Methods).

The corrected Pd diffusivities yielded an apparent activation energy $E_a$ of 1.10 ± 0.09 eV and ln ($D_0$ / $m^2\ s^{-1}$) = −35.3 ± 1.3 (Fig. 3a, Methods). This activation energy, obtained from individual 3D displacements of discrete atoms at the nanoscale interface, is much lower than approximately 2.76 eV for bulk Pd self-diffusion[42] and the 3.36–3.66 eV obtained from bulk Pd–Pt diffusion couples[43]. It instead lies close to the DFT-calculated migration barriers of 1.04 and 1.07 eV for Pd moving into an existing vacancy in the adjacent Pt layer in {100} and {111} Pd@Pt models[18], respectively, whereas barriers exceeding 5 eV were predicted for the vacancy-free exchange pathways[18]. For equilibrium bulk vacancy-mediated self-diffusion, the activation enthalpy includes both vacancy-formation and migration contributions[44]. At an interface separated from the free surface by only a few atomic layers, surface-associated vacancies could provide an alternative to bulk-like vacancy formation, as suggested by the subsurface vacancy seeding[18]. Our results are therefore consistent with an apparent activation energy governed predominantly by vacancy migration, with a reduced effective vacancy-formation contribution compared with bulk self-diffusion. In vacancy-mediated diffusion, the diffusivity scales with both the mobile-vacancy fraction and the atomic jump rate, and a low, approximately temperature-independent mobile-vacancy fraction can suppress the prefactor $D_0$[45]. Within this interpretation, the $D_0$, approximately ten orders of magnitude smaller than that of bulk Pd[42], may reflect limited time-averaged mobile-vacancy availability, compatible with surface-associated vacancy supply[18] and possible vacancy annihilation at the nearby free surface[13].

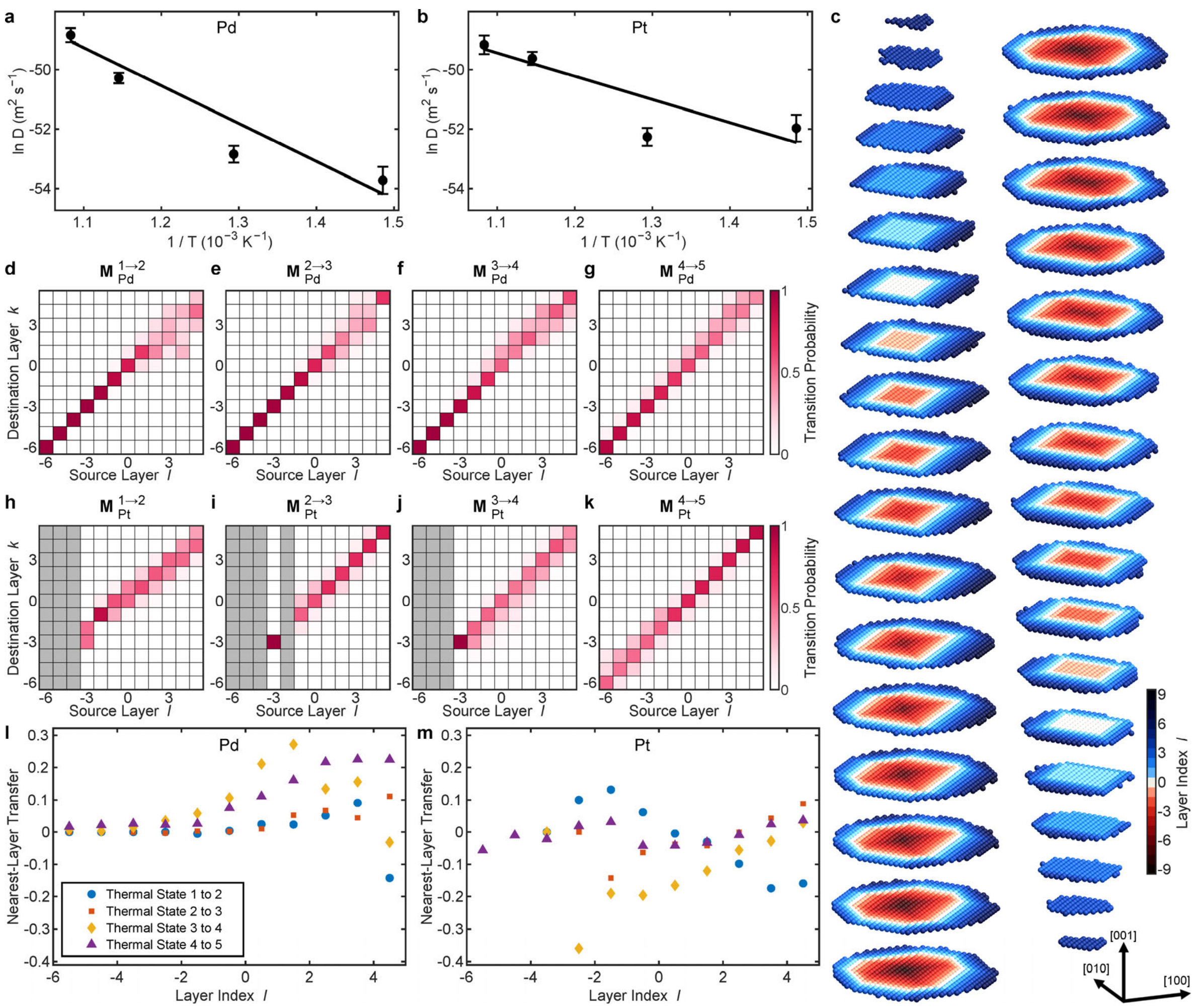

**Figure 3 | Identity-corrected 3D diffusivity and layer-resolved transport. a**,**b**, Arrhenius plots of the identity-corrected Einstein-type diffusivities for Pd (a) and Pt (b), obtained from the Jonker–Volgenant pseudo-displacements after simulation-based correction. Error bars denote standard deviations across the 200 simulation-calibration realizations, and solid lines show the corresponding Arrhenius relation from the obtained $E_a$ and $D_0$ (Methods). **c**, Atomic-layer indices used for the layer analysis; colours denote the layer index $l$, with $l$ = 0 representing the interface. **d**–**g**, Source-normalized Pd layer-transition matrices $\mathbf{M}_{\mathrm{Pd}}^{j\rightarrow j+1}$, where $j$ and $j$ + 1 denote the previous and next thermal states, respectively. **h**–**k**, Corresponding Pt matrices $\mathbf{M}_{\mathrm{Pt}}^{j\rightarrow j+1}$. The matrices were calculated directly from the Jonker–Volgenant endpoint assignments before correction. **l**,**m**, Atom-number-normalized nearest-layer transfer for Pd (l) and Pt (m) for each interval (Methods). Positive and negative values denote outward and inward transfer, respectively. Transfers between adjacent layers $i$ and $i$ + 1 are plotted at their midpoint, $l = i + 0.5$. Grey columns in (h)–(j) indicate unpopulated source layers.

Applying the same ensemble-fitting procedure to Pt gave an $E_a$ of 0.67 ± 0.10 eV and ln ($D_0$ / m² s⁻¹) = −40.8 ± 1.4 (Fig. 3b), an $E_a$ about 40% lower than the Pd value. The Pt diffusivities, however, deviate from a single Arrhenius relation, especially in the lower temperature interval. Because Pt forms the shell, the extracted Pt diffusivities are more susceptible to shell reshaping and surface diffusion, particularly at lower temperatures when buried-interface transport is weaker. Therefore, we report the Pd result as the primary measure of buried-interface transport, while the lower Pt value may retain contributions from surface-related processes and should not be interpreted as evidence of a species-dependent interfacial barrier. Using correction factors obtained from the ideal spherical core–shell model yielded Pd and Pt activation energies consistent with those from the experiment-based geometry within uncertainty, further supporting the robustness of the calibration to particle geometry (Supplementary Fig. 3j,k).

## Layer-resolved atomic transport

The corrected diffusivity measures overall 3D mobility, but the ensemble mean-squared displacement cannot identify the destination layer of each atom or directly describe chemical redistribution relative to the interface. We therefore analysed the experimental Jonker–Volgenant pseudo-trajectories directly using fixed, interface-referenced atomic-layer indices (Methods). The interface was assigned $l = 0$, with $l < 0$ toward the Pd core and $l > 0$ toward the Pt shell (Fig. 3c). Each pseudo-trajectory therefore gives an apparent same-layer, inward or outward endpoint transition. For each source layer $l$ of the thermal state $j$, the source-normalized matrix $\mathbf{M}_{kl}^{j \to j+1}$ gives the fraction of atoms assigned to destination layer $k$ in the next thermal state $j + 1$ (Fig. 3d–k, Supplementary Fig. 4, Methods).

Between thermal states 1 and 2 (Fig. 3d), the Pd matrix was nearly diagonal for layers deep within the core ($l < 0$), indicating strong same-layer endpoint retention. Off-diagonal assignments (mostly $l$ to $l \pm 1$) were concentrated within a few layers of the interface ($0 \leq l \leq 3$) and were most pronounced between thermal states 3 and 4 (Fig. 3f, Supplementary Fig. 4j), showing that the resolved Pd redistribution was localized around the buried interface. Shell-side ($l > 0$) Pt transport is the counterpart to core-side ($l < 0$) Pd transport, but the shell-side Pt matrices are broader (Fig. 3h–k, Supplementary Fig. 4k), in contrast to the nearly diagonal core-side Pd matrices (Fig. 3d–g, Supplementary Fig. 4j). This reflects additional contributions from shell reshaping and near-surface diffusion, beyond buried-interface transport.

Figure 3l,m show the normalized nearest-layer transfer of Pd and Pt, respectively. Overall, Pd moved toward the shell and Pt toward the core. These opposing transfers were strongest between thermal states 3 and 4. Between thermal states 4 and 5, the corrected diffusivity continued to increase (Fig. 3a,b), whereas net layer transfer did not increase correspondingly and even decreased relative to the preceding thermal-state interval. Progressive alloying reduced the core–shell composition gradient, so high atomic mobility was accompanied by less net compositional redistribution. The more diagonal matrices in this final interval, especially for shell-side Pt, may also partly reflect stronger minimum-distance compression during Jonker–Volgenant assignments, because the mixed configuration contains more nearby same-species candidates. These results therefore distinguish increasing 3D atomic mobility from declining net composition-changing transport as the interface becomes more mixed.

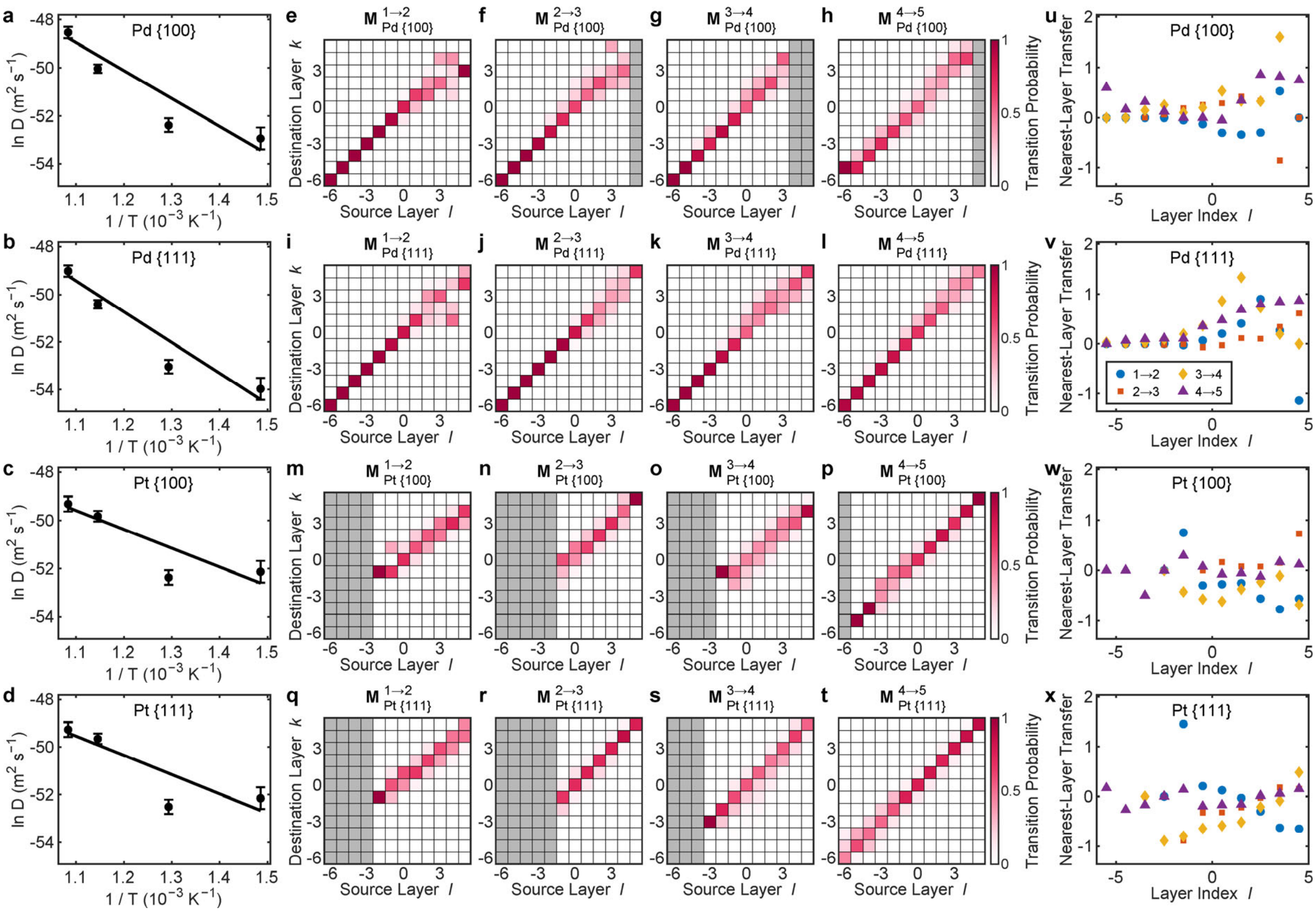

**Figure 4 | Facet-resolved 3D diffusivity and atomic-layer transport. a–d**, Arrhenius plots of the identity-corrected Einstein-type diffusivities for Pd associated with {100} (a) and {111} (b) facets and Pt associated with {100} (c) and {111} (d) facets. Error bars denote standard deviations across the 200 simulation-calibration realizations, and solid lines show the corresponding Arrhenius relation from the $E_a$ and $D_0$ (Methods). **e–l**, Source-normalized Pd layer-transition matrices for {100} (e–h) and {111} (i–l) regions over consecutive thermal-state intervals. **m–t**, Corresponding Pt matrices for {100} (m–p) and {111} (q–t) regions. The matrices were calculated directly from the uncorrected Jonker–Volgenant endpoint assignments. **u–x**, Atom-number- and cross-layer-connectivity-normalized nearest-layer transfer for Pd {100} (u), Pd {111} (v), Pt {100} (w) and Pt {111} (x). Positive and negative values denote outward and inward transfer, respectively. Transfers between adjacent layers $i$ and $i + 1$ are plotted at their midpoint, $l = i + 0.5$. Grey columns in (e)–(t) indicate unpopulated source layers.

## Facet dependence of interfacial transport

To examine crystallographic dependence, we assigned each pseudo-trajectory to a {100}- or {111}-associated sector within its layer and retained only pseudo-trajectories that remained associated with the same individual facet at both endpoints (Methods). Applying the same species- and interval-specific identity corrections to the {100}- and {111}-associated subsets gave apparent Pd activation energies of 1.00 ± 0.09 eV for {100} and 1.11 ± 0.09 eV for {111}; the corresponding Pt values were 0.67 ± 0.10 and 0.70 ± 0.10 eV (Fig. 4a–d), with no discernible facet dependence. In a fully coordinated fcc lattice, four of the twelve nearest-neighbour directions lie within a {100} plane and eight connect adjacent layers, whereas six are in-plane and six connect adjacent {111} layers. This difference changes the partition between in-plane and cross-layer transitions, but not the full nearest-neighbour jump length, whereas the Jonker–Volgenant assignment minimizes the full 3D distance without separately weighting in-plane and cross-layer motion. The full-3D Arrhenius comparison therefore requires no cross-layer connectivity normalization. The absence of a resolved facet difference is consistent with calculations showing comparable vacancy-mediated migration barriers in deeper subsurface layers beneath Pd@Pt {100} and {111} surfaces[18].

Facet-resolved interlayer transitions nevertheless revealed a distinction hidden in the scalar diffusivity (Fig. 4e–t, Supplementary Fig. 5). The source-normalized matrices, calculated directly from the Jonker–Volgenant pseudo-trajectories, showed greater same-layer retention for Pd in {111} regions, whereas {100} regions showed stronger neighbouring-layer transitions (Fig. 4e–l, Supplementary Fig. 5s). This is consistent with their six and eight cross-layer nearest-neighbour pathways for {111} and {100}, respectively. For an fcc lattice with lattice constant $a$, the corresponding interface-normal components of a nearest-neighbour jump (i.e., fcc interlayer distance) are $\frac{a}{\sqrt{3}}$ and $\frac{a}{2}$, respectively, so the pathway numbers and squared normal distance compensate exactly because $6\left(\frac{a}{\sqrt{3}}\right)^2 = 8\left(\frac{a}{2}\right)^2$. In the ideal fcc random-walk limit, the two facet families can therefore show different transition probabilities in the layer coordinate $l$, yet yield the same interface-normal mean-squared displacement once their different layer spacings are accounted for. Facet-dependent discrete layer transitions can thus coexist with isotropic displacement-based transport after coarse-graining. Normalization by the number of cross-layer pathways likewise brought the nearest-layer transfer profiles into closer agreement near the interface, particularly between thermal states 4 and 5 (Fig. 4u–x, Supplementary Fig. 5w,x; Methods). The closer agreement after normalization suggests that the facet dependence partly reflects the different numbers of available cross-layer pathways. The Pt matrices showed a weaker and less systematic facet contrast, consistent with the surface-related limitations described above. Our results therefore distinguish facet-dependent redistribution in discrete layer indices from the full 3D mobility, for which no robust facet dependence was resolved.

## Discussion

At a buried interface only a few atomic layers wide, a scalar diffusion coefficient is not a complete transport law. It averages over motion parallel and normal to the interface, different crystallographic regions, and atomic motion that does or does not change local composition. Pulse–quench AET separates these quantities through identity-corrected full-3D mobility and direct species-, layer- and facet-resolved endpoint statistics. This distinction resolves a basic ambiguity of concentration-profile measurements: slow compositional evolution can reflect either low atomic mobility or a weakened chemical driving force while atoms remain mobile. The present framework therefore provides the fundamental discrete atomic statistics from which continuum diffusion emerges.

The layer-transition matrices further extend this approach from measurement towards prediction. The matrices define empirical atomic-layer transport operators for the retained atomic populations: each source-normalized matrix $\mathbf{M}^j$ maps the layer-population vector $\mathbf{n}^j$ at thermal state $j$ to the next thermal state $j + 1$ according to $\mathbf{n}^{j+1} = \mathbf{M}^j\mathbf{n}^j$. Once their transferability is validated under repeated identical conditions, ordered products of these operators could propagate layer populations through prescribed thermal histories while retaining atom counts, chemical identity and atomic-layer resolution. Incorporating facet-retention and facet-change channels would further extend this framework to complete orientation-resolved interface evolution. These experimentally measured operators could also constrain kinetic Monte Carlo, phase-field and device-scale models, creating a multiscale route from 3D atomic configurations to predictive interface evolution without transferring bulk diffusion parameters directly to a nanometre-scale boundary.

Facet resolution reveals another fundamental point: similar full-3D mobilities do not imply identical local transport pathways. In the nanoscale fcc Pd/Pt interface, {100} and {111} regions show different discrete interlayer transport due to differences in the intralayer and cross-layer configurations of nearest-neighbour bonds, yet these differences compensate after coarse-graining. Measuring these underlying statistics offers a direct way to determine when strain, defects, segregation, finite-facet geometry or lower crystal symmetry prevent this compensation and produce genuinely anisotropic transport.

This capability is important wherever redistribution by only one or two atomic layers changes function. Diffusion and intermixing can alter strain transfer and catalytic activity in core–shell metals, band alignment and carrier dynamics in semiconductor quantum dots, and contact resistance at semiconductor interfaces[33,46–51]. The strategy can therefore be adapted to core–shell nanocrystals, multilayers, semiconductor contacts, diffusion barriers and other crystalline heterointerfaces during synthesis, thermal processing and operation.

Our work converts sequential, species-resolved 3D configurations of the same buried interface into both identity-corrected diffusivities and layer- and facet-resolved transport operators. For the nanoscale fcc Pd/Pt interfaces, this yields an apparent Pd activation energy of 1.10 ± 0.09 eV, consistent with migration-limited vacancy-mediated transport, while exposing atomic-layer and facet dependence hidden by a scalar diffusivity. More broadly, our framework transforms nanointerface diffusion from a parameter inferred from averaged profiles into experimentally resolved 3D atomic redistribution with simulation-calibrated mobility and discrete layer-resolved transport statistics. It opens an experimental route from individual atomic rearrangements towards predictive control of interface evolution, one atom and one atomic layer at a time.

## References

1. Fick, A. Ueber Diffusion. *Ann. Phys. Chem.* **94**, 59–86 (1855).
2. Onsager, L. Reciprocal relations in irreversible processes. I. *Phys. Rev.* **37**, 405–426 (1931).
3. Onsager, L. Reciprocal relations in irreversible processes. II. *Phys. Rev.* **38**, 2265–2279 (1931).
4. Einstein, A. Über die von der molekularkinetischen Theorie der Wärme geforderte Bewegung von in ruhenden Flüssigkeiten suspendierten Teilchen. *Ann. Phys.* **322**, 549–560 (1905).

5. Campbell, C. E. Diffusivity and Mobility Data. in *ASM Handbook, Volume 22A: Fundamentals of Modeling for Metals Processing* (eds David U. Furrer & S. Lee Semiatin) vol. 22 171–181 (ASM International, Materials Park, OH, 2009).
6. Smigelskas, A. D. & Kirkendall, E. O. Zinc diffusion in alpha brass. *Trans. Am. Inst. Min. Metall. Eng.* **171**, 130–142 (1947).
7. Darken, L. S. Diffusion, mobility and their interrelation through free energy in binary metallic systems. *Trans. Am. Inst. Min. Metall. Eng.* **175**, 184–201 (1948).
8. van Dal, M. J. H., Gusak, A. M., Cserháti, C., Kodentsov, A. A. & van Loo, F. J. J. Microstructural stability of the Kirkendall plane in solid-state diffusion. *Phys. Rev. Lett.* **86**, 3352–3355 (2001).
9. Mehrer, H. *Diffusion in Solids*. vol. 155 (Springer, Berlin, Heidelberg, 2007).
10. Van der Ven, A., Yu, H.-C., Ceder, G. & Thornton, K. Vacancy mediated substitutional diffusion in binary crystalline solids. *Prog. Mater. Sci.* **55**, 61–105 (2010).
11. Erdélyi, Z. & Beke, D. L. Nanoscale volume diffusion. *J. Mater. Sci.* **46**, 6465–6483 (2011).
12. Balogh, Z. *et al.* Transition from anomalous kinetics toward Fickian diffusion for Si dissolution into amorphous Ge. *Appl. Phys. Lett.* **92**, 143104 (2008).
13. Chee, S. W. *et al.* Interface-mediated Kirkendall effect and nanoscale void migration in bimetallic nanoparticles during interdiffusion. *Nat. Commun.* **10**, 2831 (2019).
14. Yin, Y. *et al.* Formation of Hollow Nanocrystals Through the Nanoscale Kirkendall Effect. *Science* **304**, 711–714 (2004).
15. Gärtner, D. *et al.* Techniques of tracer diffusion measurements in metals, alloys and compounds. *Diffus. Found.* **29**, 31–73 (2021).
16. Ehrlich, G. & Hudda, F. G. Atomic view of surface self-diffusion: tungsten on tungsten. *J. Chem. Phys.* **44**, 1039–1049 (1966).
17. Swartzentruber, B. S. Direct measurement of surface diffusion using atom-tracking scanning tunneling microscopy. *Phys. Rev. Lett.* **76**, 459–462 (1996).
18. Vara, M. *et al.* Understanding the thermal stability of palladium–platinum core–shell nanocrystals by in situ transmission electron microscopy and density functional theory. *ACS Nano* **11**, 4571–4581 (2017).
19. Schweizer, P. *et al.* Atomic scale volume and grain boundary diffusion elucidated by in situ STEM. *Nat. Commun.* **14**, 7601 (2023).
20. Futazuka, T. *et al.* Direct observation of substitutional and interstitial dopant diffusion in oxide grain boundary. *Nat. Commun.* **16**, 9043 (2025).
21. Ishikawa, R. *et al.* Real-time tracking of three-dimensional atomic dynamics of Pt trimer on TiO2 (110). *Sci. Adv.* **10**, eadk6501 (2024).
22. Ishikawa, R. *et al.* 3D dynamic structure of a Pt nanoparticle on SrTiO3 (001) during in-situ heating atomic-resolution ADF STEM imaging. *Nat. Commun.* **17**, 1860 (2026).
23. Skorikov, A. *et al.* Quantitative 3D Characterization of Elemental Diffusion Dynamics in Individual Ag@Au Nanoparticles with Different Shapes. *ACS Nano* **13**, 13421–13429 (2019).
24. Albrecht, W., Van Aert, S. & Bals, S. Three-Dimensional Nanoparticle Transformations Captured by an Electron Microscope. *Acc. Chem. Res.* **54**, 1189–1199 (2021).
25. Chen, Q. *et al.* Estimation of Temperature Homogeneity in MEMS-Based Heating Nanochips via Quantitative HAADF-STEM Tomography. *Part. Part. Syst. Charact.* **41**, 2300070 (2024).
26. Albrecht, W. *et al.* Thermal Stability of Gold/Palladium Octopods Studied in Situ in 3D: Understanding Design Rules for Thermally Stable Metal Nanoparticles. *ACS Nano* **13**, 6522–6530 (2019).
27. Craig, T. M. *et al.* Continuous three-dimensional imaging of nanoscale dynamics by in situ electron tomography. Preprint at https://doi.org/10.48550/arXiv.2603.29462 (2026).
28. Vanrompay, H. *et al.* 3D characterization of heat-induced morphological changes of Au nanostars by fast *in situ* electron tomography. *Nanoscale* **10**, 22792–22801 (2018).
29. Chi, M. *et al.* Surface faceting and elemental diffusion behaviour at atomic scale for alloy nanoparticles during in situ annealing. *Nat. Commun.* **6**, 8925 (2015).
30. Xu, R. *et al.* Three-dimensional coordinates of individual atoms in materials revealed by electron tomography. *Nat. Mater.* **14**, 1099–1103 (2015).

31. Yang, Y. *et al.* Deciphering chemical order/disorder and material properties at the single-atom level. *Nature* **542**, 75–79 (2017).
32. Zhou, J. *et al.* Observing crystal nucleation in four dimensions using atomic electron tomography. *Nature* **570**, 500–503 (2019).
33. Jo, H. *et al.* Direct strain correlations at the single-atom level in three-dimensional core-shell interface structures. *Nat. Commun.* **13**, 5957 (2022).
34. Li, Z. *et al.* Probing the atomically diffuse interfaces in Pd@Pt core-shell nanoparticles in three dimensions. *Nat. Commun.* **14**, 2934 (2023).
35. Xie, J. *et al.* Tracking atomic-scale interdiffusion in immiscible bimetallic nanoparticles via four-dimensional electron tomography. Preprint at https://doi.org/10.48550/arXiv.2606.12150 (2026).
36. Jonker, R. & Volgenant, A. A shortest augmenting path algorithm for dense and sparse linear assignment problems. *Computing* **38**, 325–340 (1987).
37. van Omme, J. T., Zakhozheva, M., Spruit, R. G., Sholkina, M. & Pérez Garza, H. H. Advanced microheater for *in situ* transmission electron microscopy; enabling unexplored analytical studies and extreme spatial stability. *Ultramicroscopy* **192**, 14–20 (2018).
38. Miao, J., Ercius, P. & Billinge, S. J. L. Atomic electron tomography: 3D structures without crystals. *Science* **353**, aaf2157 (2016).
39. Zhou, J., Yang, Y., Ercius, P. & Miao, J. Atomic electron tomography in three and four dimensions. *MRS Bull.* **45**, 290–297 (2020).
40. Tang, M. *et al.* Pd–Pt nanoalloy transformation pathways at the atomic scale. *Mater. Today Nano* **1**, 41–46 (2018).
41. Yun, K. *et al.* Monte Carlo simulations of the structure of Pt-based bimetallic nanoparticles. *Acta Mater.* **60**, 4908–4916 (2012).
42. Peterson, N. L. Isotope Effect in Self-Diffusion in Palladium. *Phys. Rev.* **136**, A568–A574 (1964).
43. Baheti, V. A., Ravi, R. & Paul, A. Interdiffusion study in the Pd–Pt system. *J. Mater. Sci. Mater. Electron.* **24**, 2833–2838 (2013).
44. Schaefer, H.-E. & Banhart, F. Thermal equilibrium vacancies in platinum studied by positron annihilation. *Phys. Status Solidi A* **104**, 263–272 (1987).
45. Mantina, M. *et al.* First-Principles Calculation of Self-Diffusion Coefficients. *Phys. Rev. Lett.* **100**, 215901 (2008).
46. Zhang, J. *et al.* Platinum Monolayer Electrocatalysts for O2 Reduction: Pt Monolayer on Pd(111) and on Carbon-Supported Pd Nanoparticles. *J. Phys. Chem. B* **108**, 10955–10964 (2004).
47. Xie, S. *et al.* Atomic Layer-by-Layer Deposition of Pt on Pd Nanocubes for Catalysts with Enhanced Activity and Durability toward Oxygen Reduction. *Nano Lett.* **14**, 3570–3576 (2014).
48. Wang, X. *et al.* Palladium–platinum core-shell icosahedra with substantially enhanced activity and durability towards oxygen reduction. *Nat. Commun.* **6**, 7594 (2015).
49. Jing, L. *et al.* Insight into Strain Effects on Band Alignment Shifts, Carrier Localization and Recombination Kinetics in CdTe/CdS Core/Shell Quantum Dots. *J. Am. Chem. Soc.* **137**, 2073–2084 (2015).
50. Beane, G. A., Gong, K. & Kelley, D. F. Auger and Carrier Trapping Dynamics in Core/Shell Quantum Dots Having Sharp and Alloyed Interfaces. *ACS Nano* **10**, 3755–3765 (2016).
51. Song, Q., Zhou, J. & Chen, G. Significant reduction in semiconductor interface resistance via interfacial atomic mixing. *Phys. Rev. B* **105**, 195306 (2022).

## Methods

### Sample preparation

*Chemicals and materials*

Carbon black (Vulcan XC-72R) was purchased from Cabot Corporation. Palladium(II) chloride solution ($H_2PdCl_4$, 10 wt.% Pd in HCl solution) and $K_2PtCl_4$ powder (98%) were purchased from Lab Network. Sodium hydroxide (NaOH, 97%) was purchased from Sigma-Aldrich. Carbon monoxide gas (99.998%) was supplied from Air Liquide Korea.

*Synthesis of Pd@Pt core–shell nanoparticles*

To synthesize Pd@Pt core–shell nanoparticles on carbon, Pd/C nanoparticles were first synthesized. 3.5 g carbon black was dispersed into 1.5 L of distilled water in a three-neck round-bottom flask using a high shear mixer for 10 min at 6000 rpm. Afterwards, the flask was transferred to a bath-type 40 kHz sonicator and sonicated for extra 10 min to achieve uniform dispersion. After inserting a magnetic stirring bar and transferring the flask to a magnetically stirred reactor, 15 mL of Pd(II) chloride solution was added to the mixture while purging the solution with $N_2$ gas (1000 mL/min) for 20 min. After adjusting the pH to 11 by slowly adding 1.0 M NaOH to the mixture, CO gas (1000 mL/min) was bubbled to the solution for 30 min to induce reduction of Pd(II) chloride. After the reaction, Pd/C nanoparticles were filtered and dried in a vacuum oven for 20 h at 70 °C.

After Pd/C synthesis, Pd@Pt core–shell nanoparticles were synthesized by CO-Adsorption-Induced Deposition (CO-AID) method[52]. 1 g of the synthesized Pd/C nanoparticles was dispersed into 1 L of distilled water in a three-neck round-bottom flask using a high shear mixer for 10 min at 6000 rpm. The mixture was further sonicated in a bath-type 40 kHz sonicator for 10 min. After inserting a magnetic stirring bar and transferring the flask to a magnetically stirred reactor, the mixture was purged with $N_2$ gas (1000 mL/min) for 10 min to ensure an inert atmosphere and removal of impurities. Next, CO gas (1000 mL/min) was introduced to the mixture for 10 min to induce adsorption of CO on Pd surfaces. After CO purging, $N_2$ gas was purged again to the mixture for 10 min to eliminate CO gas dissolved in water, so that the CO molecules only exist on Pd surfaces. $K_2PtCl_4$ solution (3 mL, 0.1 $g_{Pt}$/mL $H_2O$) was then added to the mixture to form the Pt shells on Pd nanoparticles. After waiting for 5 min, the mixture was filtered to obtain the resultant Pd@Pt/C nanoparticles, and the product was dried in a vacuum oven for 20 h at 70 °C. Since each step of CO-AID creates one monolayer shell, the product was re-dispersed into distilled water and treated with the same procedure to create another Pt layer on the surface. Three consecutive CO-AID cycles yielded Pd@Pt/C 3ML nanoparticles.

### Pulse–quench ADF-STEM heating screening

Electron microscopy specimens were prepared by depositing an ethanol dispersion of synthesized Pd@Pt nanoparticles onto 50-nm-thick silicon nitride membranes on MEMS heating chips compatible with the DENSsolutions Wildfire tomography heating holder. To identify a temperature–time window in which interfacial diffusion could be activated while preserving the nanoparticle structure, we performed 2D ADF-STEM heating screening using a pulse–quench protocol. In this screening experiment, nanoparticles were heated to defined temperatures for controlled dwell times, rapidly quenched to room temperature and then imaged by ADF-STEM. This procedure allowed us to evaluate thermally induced structural and chemical evolution before selecting the conditions used for pulse–quench AET measurements (Supplementary Fig. 1).

For the temperature-dependent analysis, three Pd@Pt nanoparticles on a single MEMS heating chip were selected and imaged by ADF-STEM at room temperature before heating. The chip was then subjected to repeated heat–hold–quench–image cycles. At each target temperature of 200, 300, 400, 500 and 600 °C, the chip was heated and held for 5 min, quenched to room temperature and imaged by ADF-STEM. This cycle was repeated three times at each temperature, giving accumulated heating times of 5, 10 and 15 min.

For the time-dependent analysis, four Pd@Pt nanoparticles on a different MEMS heating chip were selected and first imaged at room temperature. The chip was then repeatedly heated to 500 °C, held for defined durations, quenched to room temperature and imaged by ADF-STEM. The initial sequence used 2-min heating increments, giving accumulated heating times from 2 to 30 min in 2-min steps. The same heat–hold–quench–image procedure was then continued using 10-min increments, giving accumulated heating times of 40, 50 and 60 min at 500 °C. Finally, on the same chip, additional pulse–quench measurements were performed after heating at 600 °C for 10 min, 700 °C for 10 min and 800 °C for 5 min.

ADF-STEM images were acquired using a double Cs-corrected Titan Cubed G2 60–300 microscope (FEI) operated at an accelerating voltage of 300 kV. Imaging was performed in ADF-STEM mode with a convergence semi-angle of 25.2 mrad, detector collection range of 38–200 mrad and a beam current of 50 pA. Each image was recorded with 1024 × 1024 pixels, a pixel size of 35.8 pm, and a dwell time of 4.00 μs per pixel.

**Tomography data acquisition**

A Pd@Pt nanoparticle specimen similarly prepared on a MEMS heating chip was mounted on a DENSsolutions Wildfire tomography heating holder and heated to 300 °C for 900 s inside the microscope. The specimen was rapidly heated to the target temperature under high vacuum, quenched to room temperature within a few seconds and then measured by AET; the full temperature–time profiles for all five thermal cycles are shown in Supplementary Fig. 2. After quenching, an ADF-STEM tomographic tilt series was acquired from a selected Pd@Pt nanoparticle, corresponding to thermal state 1. The same chip was subsequently heated to 400 °C for 900 s, quenched to room temperature and measured in the same manner to obtain the tomographic tilt series for thermal state 2 from the same nanoparticle. The procedure was repeated after 900-s heating at 500, 600 and 650 °C to obtain thermal states 3, 4 and 5, respectively (Supplementary Fig. 6–10).

All images were acquired in ADF-STEM mode at an acceleration voltage of 300 kV and a beam current of 15 pA. To correct for stage drift during data acquisition, three consecutive 1024 × 1024 pixel images were collected at each tilt angle with a dwell time of 3 μs per pixel. Detailed microscope parameters for tomography data acquisition are provided in Supplementary Table 1. Each tilt series was acquired with a total electron dose ranging from $2.2 \times 10^5$ $e$ Å$^{-2}$ to $2.8 \times 10^5$ $e$ Å$^{-2}$. At an accelerating voltage of 300 kV, the maximum kinetic energies transferable to initially stationary Pd and Pt atoms are approximately 8.0 and 4.4 eV, respectively[53], well below reported bulk displacement thresholds of approximately 34 eV for Pd and 33 eV for Pt[54,55]. Direct bulk knock-on defect production is therefore not expected, although this comparison does not exclude beam-assisted surface processes. To verify that electron-beam-induced structural changes were negligible, zero-degree projections were recorded three times for each tilt series, at the beginning, middle and end of the acquisition. No substantial structural changes were observed in these repeated projections (Supplementary Fig. 11), indicating that the nanoparticle remained stable within the applied dose range.

**Image post-processing**

All tilt series were post-processed using established AET procedures, including drift correction, scan-distortion correction, image denoising, background subtraction and tilt-series alignment[30,31,33,56–61].

(I) Drift and scan-distortion correction: For each tilt angle, linear stage drift was estimated from the three consecutively acquired ADF-STEM images and corrected using an affine transformation. Scan distortion was then corrected using a scan-distortion matrix calibrated from images of a single-crystal Si(110) standard sample. After drift and scan-distortion correction, the three consecutive images were averaged to generate one experimental projection image for each tilt angle.

(II) Image denoising: To reduce mixed Gaussian–Poisson noise in the ADF-STEM images, we applied the block-matching and 3D filtering (BM3D) algorithm[62]. The noise parameters were estimated directly from the experimental tilt series using the statistics of the three consecutively acquired images at each tilt angle.

(III) Background subtraction and tilt-series alignment: To remove the background signal, a 2D mask slightly larger than the nanoparticle boundary was defined for each projection image. The background inside the mask was estimated by solving the Dirichlet boundary-value problem for the discrete Laplace equation and was subtracted from the denoised image. Each tilt series was then aligned with sub-pixel accuracy using a combination of centre-of-mass[63], common-line[31] and cross-correlation[64,65] alignment methods.

**RESIRE reconstruction**

After image post-processing, 3D tomograms were reconstructed from the processed tilt series using the RESIRE algorithm[66]. For tilt series that showed improvement after refinement, reconstruction-based in-plane rotational and translational realignment was additionally performed[56]. Final 3D tomograms were then reconstructed using 1,000 RESIRE iterations with an oversampling ratio of 4; detailed reconstruction parameters are summarized in Supplementary Table 1.

**Atom tracing, lattice assignment and chemical species classification**

The 3D atomic coordinates of all atoms in the nanoparticle were first determined by fitting a 3D Gaussian function to a 5 × 5 × 5 voxel volume centred on each local maximum of the 3D tomograms, with a minimum interatomic-distance constraint of 2.0 Å[30–33,39,56–58,67,68]. To identify atomic positions that may have been missed because of atom elongation from the missing wedge or minor reconstruction imperfections, the same 3D Gaussian fitting process was additionally applied to 2D local maxima identified from the tomogram slices along the fcc [001] direction[33]. Experimental 3D atomic positions were assigned to ideal fcc lattice sites using the iterative lattice-fitting procedure below[33,58]:

(a) The atom closest to the centroid of all reconstructed atomic coordinates was selected as the origin of an initial ideal fcc lattice.

(b) Candidate fcc sites were then generated from this origin using the current fcc lattice vectors. In the first iteration, these lattice vectors were initialized with a lattice constant of 3.9 Å. For each generated lattice site, if a traced atom was located within 25% of the fcc nearest-neighbour distance, the atom was assigned to that lattice site.

(c) The nearest-neighbour search was then propagated from all newly assigned fcc sites to generate additional neighbouring lattice sites, and the assignment procedure was repeated until no further atoms could be assigned.

(d) The fcc lattice vectors were refined by optimizing the translation, 3D rotation and lattice constant to minimize the discrepancy between the experimental atomic positions and their corresponding ideal lattice sites.

(e) Steps (a)–(d) were repeated independently for each thermal state until the fitted lattice vectors converged. The arithmetic mean of the fitted lattice constants was used as the common effective lattice constant in the subsequent analyses.

Additional tracing was then performed using the fitted fcc lattice as a guide; if no atom was present within a radius corresponding to one-third of the nearest-neighbour distance from a fitted lattice position, an atomic position was assigned by fitting a 3D Gaussian function to the 5 × 5 × 5 voxel tomogram intensity around that lattice position. If this fitting procedure did not converge, local lattice fitting was performed using neighbouring atomic positions, and the fitted lattice centre was added as the atomic position.

Because the nanoparticle was isolated and showed no visible change in size during the pulse–quench heating experiment, the surface boundaries were adjusted to conserve the total number of traced atoms across thermal states. An alpha-shape boundary based on Delaunay triangulation[69] was applied to the atomic model of each thermal state, with the shrink factor chosen such that the number of atoms enclosed by the boundary was as close as possible to that of the thermal state containing the fewest traced atoms. The model for each thermal state was then obtained by removing atoms in descending order of their distance from the geometric centroid of all traced atomic coordinates until all thermal-state models contained the same total number of atoms. The fitted lattices of these resulting models were obtained by the lattice-fitting procedure above.

For the atomic model of each thermal state, the traced atoms were first classified as Pd or Pt using a k-means-clustering-based procedure[31–33]. The intensity within a 3 × 3 × 3-voxel tomographic volume centred on each atom was integrated, and the atoms were initially classified as Pd or Pt based on a threshold defined as the mean of the integrated intensities of all traced atoms in that thermal state. Next, 7 × 7 × 7-voxel volumes centred on the atoms assigned as Pd and Pt were separately averaged to obtain species-specific reference volumes. We denote the intensity of the *v*th voxel in the averaged Pd and Pt reference volumes for the *j*th thermal state as $A^j_{\mathrm{Pd},v}$ and $A^j_{\mathrm{Pt},v}$, respectively. For the *i*th atom, with voxel intensity $I^j_{iv}$, the errors relative to the Pd and Pt reference volumes, $E^j_{i,\mathrm{Pd}}$ and $E^j_{i,\mathrm{Pt}}$, were calculated as

$$E^j_{i,\mathrm{Pd}} = \sum_v \left| I^j_{iv} - A^j_{\mathrm{Pd},v} \right|, \quad E^j_{i,\mathrm{Pt}} = \sum_v \left| I^j_{iv} - A^j_{\mathrm{Pt},v} \right|.$$

Each atom was then reassigned to the species yielding the smaller error. The Pd and Pt reference volumes were recalculated from the updated assignments, and the classification was iterated until the species assignments converged.

Because the numbers of Pd and Pt atoms should be conserved during diffusion, a constrained version of the clustering procedure was further applied. The same iterative procedure described above was used, except that the number of Pd atoms was fixed as $N_{\mathrm{Pd}}$. At each iteration, the error metric $E^j_{i,\mathrm{Pd}} - E^j_{i,\mathrm{Pt}}$ was calculated for all atoms in this case, and the $N_{\mathrm{Pd}}$ atoms with the lowest errors were assigned as Pd, with the remainder assigned as Pt. The reference volumes were then recalculated from the updated assignments, and the procedure was repeated until convergence. The same $N_{\mathrm{Pd}}$ was imposed on the constrained clustering of all thermal-state atomic models. The final $N_{\mathrm{Pd}}$ was chosen to minimize the total discrepancy between the constrained and unconstrained clustering results across all thermal states. Because the total number of atoms was identical for every thermal state, this constraint also ensured conservation of the Pt-atom number.

To establish a common lattice-coordinate system across the thermal states, the independently fitted lattices were aligned sequentially. For each pair of consecutive thermal states, integer translations along the three primitive fcc lattice vectors were examined, and the translation maximizing the number of identical species assignments among lattice sites occupied in both states was selected.

Following lattice alignment, the constrained classification procedure described above was repeated with an additional constraint on the centre of mass of the internal Pd atoms as follows. First, the unconstrained assignments were transferred from the traced atomic positions to their corresponding fitted lattice sites. Outermost surface atoms were identified based on the alpha-shape algorithm with a shrink factor of 1, and these atoms and the atoms in their nearest-neighbour layer were excluded when calculating the centre of mass. For each thermal state, the centre of mass of the remaining internal Pd atoms was calculated in the common aligned lattice-coordinate system.

Using the error metrics and Pd/Pt reference volumes defined above, we further defined $z_i^j$ as the binary species assignment of atom $i$ in thermal state $j$, with $z_i^j = 1$ for Pd and $z_i^j = 0$ for Pt. For a given Pd-atom number $N_{\mathrm{Pd}}$, the assignments were obtained by minimizing

$$\min_{\mathbf{z}} \sum_j \sum_i \left[ z_i^j E_{i,\mathrm{Pd}}^j + \left(1 - z_i^j\right) E_{i,\mathrm{Pt}}^j \right]$$

subject to a fixed $N_{\mathrm{Pd}}$ in every thermal state. In addition, each Cartesian component of the centre of mass of the internal Pd atoms was constrained to lie within the tolerance of 0.001 lattice-constant-normalized coordinate units from the mean of the centres of mass of Pd atoms (based on the unconstrained classification result) of each thermal state. The Pd and Pt reference volumes were recalculated from the resulting assignments, and the optimization was repeated until convergence. At each iteration, the Pd/Pt assignments for all five thermal states were determined jointly by binary integer linear optimization using Gurobi Optimizer[70] (version 13.0.1; Gurobi Optimization, LLC). If the assignments entered a repeating cycle instead of converging, the assignment within the cycle having the fewest differences from the unconstrained classifications, summed over all thermal states, was retained. The final $N_{\mathrm{Pd}}$ was chosen to minimize the total discrepancy between the constrained and unconstrained clustering results across all thermal states. As a sensitivity test, the classification procedure was repeated using a tolerance parameter of 0.01, which changed the resulting Pd and Pt atom counts by only 8 atoms each out of 12,795 total atoms. The resulting numbers of final Pd and Pt atoms are summarized in Supplementary Table 1.

**Precision estimation using STEM multislice simulation**

Precision of the obtained 3D atomic coordinates was estimated via multislice simulation. For each thermal state, ADF-STEM projection images were generated from the experimentally determined 3D atomic model at the experimental tilt angles using multislice simulation[71–74]. The simulations were performed with a slice thickness of 2 Å, 8 frozen-phonon configurations, and aberration coefficients of 130 nm for $C_3$ and 5 mm for $C_5$. Other microscope parameters, including the acceleration voltage, convergence semi-angle and detector inner and outer collection angles, were set to match the experimental acquisition conditions listed in Supplementary Table 1. To account for the finite electron-probe size and other incoherent broadening effects, each multislice-simulated image was convolved with a Gaussian kernel with the standard deviation listed in Supplementary Table 1.

The simulated tilt series were reconstructed using the RESIRE algorithm[66] with the same reconstruction parameters used for the experimental data. Atomic coordinates were then determined from the reconstructed tomograms using the same atom-tracing and species-classification procedures described above.

To estimate coordinate precision, the atomic structures recovered from the multislice-simulated tilt series were compared with the corresponding input experimental atomic models. Atom pairs separated by less than half of the first nearest-neighbour distance in the fitted ideal fcc lattice, corresponding to 1.4 Å, were classified as common atom pairs. On average, 97.5 ± 0.7% of atoms were successfully recovered from the simulated reconstructions. The root-mean-square deviation of the common atom pairs, used as the coordinate precision estimate[30–32], was 26.2 ± 3.1 pm across all thermal states (Supplementary Table 1).

**Definition of the core–shell interface and assignment of atomic layers and facets**

The final fitted fcc lattice of thermal state 2, which showed the clearest core–shell boundary, was used to define the interface; in thermal state 1, part of the Pd core was not covered by the Pt shell. The core region was determined based on the distribution of Pd atoms, excluding the outward or isolated Pd atoms as below. First, the centroid of all Pd lattice sites was used as the origin. For each Pd site, a radial axis was drawn from this origin through the site. Pt lattice sites lying within 15° of the axis and within a perpendicular distance of five nearest-neighbour spacings from the axis were projected onto it, and their mean projected distance from the origin was calculated. The Pd site used to define the axis was classified as non-core if its distance from the origin exceeded the mean projected distance of the selected Pt atoms along that axis. Second, Pd atoms having no more than two other Pd atoms within their second-nearest-neighbour distance were considered isolated and were not included in the core definition. The lattice sites occupied by remaining Pd atoms were used as the initial representation of the core region.

To obtain a continuous, well-defined core boundary, two-dimensional slices were taken along each of the three primitive fcc lattice coordinates. In each slice, each connected component of non-core sites enclosed by core sites was reassigned to the core. The core mask was then refined by one dilation–erosion sequence followed by one erosion–dilation sequence using a lattice-neighbourhood kernel defined by integer coordinate offsets $(\Delta u, \Delta v, \Delta w)$ satisfying $\Delta u^2 + \Delta v^2 + \Delta w^2 \leq 4$. These operations filled narrow gaps and indentations and removed isolated sites, thin connections and surface protrusions. Interface sites were defined from the alpha-shape boundary of the refined core region using a shrink factor of 0.995.

To assign each interface lattice site to a facet in the {100} or {111} family, a local triangular mesh was constructed using the site and its nearest-neighbour sites in the core region, and its local normal vector was calculated from this mesh using a discrete Laplace–Beltrami operator with cotangent discretization[75]. The normal-vector field was then smoothed with a Gaussian kernel having a standard deviation of two nearest-neighbour distances. Each smoothed normal vector was compared with unit vectors along the six symmetry-equivalent <100> and eight <111> directions, and the direction with the largest dot product determined the facet assignment.

For each of the 14 facets, the fcc atomic plane parallel to that facet and containing the largest number of interface lattice sites was selected as its base plane. Together, these 14 base planes defined the $l = 0$ layer, and this layer was used as a reference for layer index and the facet assignment of every lattice site in each thermal state. For the *i*th lattice site at position $\mathbf{r}_i$, its layer index relative to base plane *p* was calculated as

$$l_i = \max_{p=1,\ldots,14} \left[ \text{round} \left( \frac{\mathbf{n}_p \cdot \mathbf{r}_i - b_p}{s_p} \right) \right],$$

where $\mathbf{n}_p$ is the outward unit normal of base plane *p*, $b_p$ is its distance from the origin, and $s_p$ is the spacing between successive occupied fcc atomic layers: $a$ / 2 for $\{100\}$ and $a$ /$\sqrt{3}$ for $\{111\}$, where $a$ is the lattice constant. Each atom was assigned to the facet whose base plane yielded the largest $l_i$. If multiple facets yielded the same maximum, the exact facet was left unresolved; if all tied facets belonged to the same $\{100\}$ or $\{111\}$ family, the atom was assigned only to that facet family.

**Lattice-exchange simulations and selection of simulation snapshots**

A starting configuration was constructed from the fitted fcc lattice of experimental thermal state 2, which showed the clearest core–shell boundary. All sites within the regularized core were initially assigned as Pd, and the remaining sites as Pt. The outermost surface sites were identified using the alpha-shape algorithm with a shrink factor of 1. At these sites and their inward nearest-neighbour sites, the experimental Pd/Pt assignments from thermal state 2 were retained. To match the experimental Pd/Pt count exactly, the 15 Pd-assigned core sites closest to the interface were reassigned as Pt.

At each simulation step, the atoms occupying a randomly selected pair of nearest-neighbour sites were exchanged. To limit the influence of surface Pd redistribution on the buried-interface calibration, exchange moves transferring Pd from outermost surface sites into the particle interior were prohibited, yielding configurations more consistent with those observed experimentally. Twenty independent simulation trajectories, each containing 1,000,000 sequential exchange steps, were generated.

To select simulation snapshots corresponding to the five experimental thermal states, layer indices were assigned to the simulated lattice sites using the procedure described above. Because the analysis focused on diffusion at the buried core–shell interface rather than surface or subsurface transport, the outermost surface sites and their nearest-neighbour sites were excluded from both the experimental and simulated configurations. For each experimental thermal state, the Pd atom count $N_{\text{Pd}}(l)$ was determined in every available layer within $l \geq 2$, where interfacial Pd diffusion was most clearly resolved. The corresponding $N_{\text{Pd}}(l)$ were calculated for every snapshot along each simulation trajectory and compared with the experimental values using the root-mean-square difference. For each thermal state, the snapshot with the minimum difference was selected independently. In every simulation trajectory, the five selected snapshots occurred at strictly increasing step numbers from thermal states 1 to 5, consistent with the experimental sequence. These ordered, identity-labelled simulation snapshots were used for the subsequent atom-matching and correction factor analyses.

To examine the dependence of the simulation results on particle geometry, we repeated the simulations using an ideal spherical core–shell model. An fcc sphere was generated with a total number of lattice sites closest to the atom count in experimental thermal state 2. The outermost surface and first subsurface sites were identified as in the experiment-based model, and their Pd/Pt identities were assigned randomly while matching the number of Pd atoms in the corresponding region of the experiment-based model. The radius of the spherical Pd core was then chosen to match the experimental Pd and Pt populations as closely as possible. The resulting model contained 62 more Pd atoms and 18 fewer Pt atoms than thermal state 2, giving 44 more atoms in total. The same nearest-neighbour exchange procedure and surface-Pd accumulation constraint were applied to generate 20 independent trajectories. For each trajectory, snapshots were taken at the same five exchange-step indices selected for the corresponding experiment-based simulation.

**Simulation-based correction factor determination for displacement calibration**

For each pair of simulation snapshots corresponding to consecutive experimental thermal states, the retained atom-identity labels allowed individual Pd and Pt atoms to be tracked between snapshots. The difference between the two lattice-site positions of each atom was defined as its labelled ground-truth 3D displacement. Note that outermost surface sites were identified using the alpha-shape algorithm with a shrink factor of 1, and atoms occupying these sites or their nearest-neighbour sites at either endpoint were excluded from the ground-truth.

Next, to mimic the experimental analysis, we introduced chemical-classification uncertainty into the simulated snapshots and then performed atom matching without using the retained atom-identity labels. First, 5% of the Pd/Pt chemical labels in each snapshot were randomly perturbed, representing the approximately 5% uncertainty in AET chemical classification for binary-alloy nanoparticles[32]. Ten independent chemical-label-perturbed realizations were generated for each of the 20 simulation trajectories.

Pd and Pt atoms were then matched separately between consecutive snapshots using the Jonker–Volgenant algorithm[36], which finds the global one-to-one assignment that minimizes the total squared Euclidean distance between occupied lattice sites. If the number of atoms assigned to a species differed between snapshots, excess atoms in the larger set were left unmatched and excluded. Also, the atoms occupying surface sites or their nearest-neighbour sites were excluded from the matching. The resulting endpoint correspondences were defined as pseudo-trajectories, and the differences between their endpoint positions as pseudo-displacements.

For each displacement set, the corresponding Einstein-type diffusivity was calculated as $D = \frac{\langle \|\Delta\mathbf{r}\|^2 \rangle}{6\Delta t}$, where $\langle \|\Delta\mathbf{r}\|^2 \rangle$ is the mean squared magnitude of the ground-truth or pseudo-displacements and $\Delta t =$

900 s. For each species and thermal-state interval, the multiplicative correction factor was defined as $C = \frac{D_{\text{labelled}}^{\text{sim}}}{D_{\text{pseudo}}^{\text{sim}}}$, where $D_{\text{labelled}}^{\text{sim}}$ and $D_{\text{pseudo}}^{\text{sim}}$ were calculated from the ground-truth and pseudo-displacements, respectively, for the same pair of simulated snapshots. One complete correction factor set was obtained for each combination of 20 simulation trajectories and 10 chemical-label perturbations, yielding 200 sets. Each set was retained separately for the subsequent Arrhenius analysis, and the arithmetic mean and sample standard deviation were calculated for each species and interval (Supplementary Fig. 3i).

**Experimental atom matching and removal of spatially correlated assignment artefacts**

Experimental pseudo-trajectories were obtained by matching Pd and Pt atoms separately between each pair of consecutive thermal states using the Jonker–Volgenant algorithm. The final chemical classifications and aligned fitted lattice-site positions were used, following the same procedure applied to the simulated snapshots.

Slight changes in surface shape caused by surface diffusion can generate spatially correlated matching artefacts. In such cases, multiple short nearest-neighbour displacement vectors connect head-to-tail to form several adjacent chains, mimicking atomic migration between different surface or interfacial regions. To identify these clusters, each maximal path or cycle of head-to-tail-connected fcc nearest-neighbour displacement vectors containing at least three distinct lattice sites was defined as a chain. Each vector in these chains was represented by its midpoint, and its local coordination was determined by counting other vector midpoints within one fcc nearest-neighbour distance. Vectors with more than two such neighbours were classified as locally clustered, indicating that multiple chains were spatially adjacent rather than forming an isolated single-chain sequence. A spatially correlated cluster was then defined as a maximal connected set of vectors in which every vector had at least one connection to a vector from a different chain within the same cluster; the cluster size was defined as the number of vectors in the set.

This procedure was applied separately to Pd and Pt for all four thermal-state intervals in all 200 simulation and chemical-label-perturbation realizations used to determine the correction factors. The largest cluster observed across these simulated cases was used as a common cluster-size cutoff. Experimental clusters larger than this cutoff (49 vectors) were classified as reshaping-related assignment artefacts, and all pseudo-trajectories belonging to these clusters were excluded from subsequent analyses, where the excluded clusters had sizes varying from 69 to 783.

**Corrected diffusion coefficients and Arrhenius analysis**

An Einstein-type pseudo-diffusivity was calculated for each species and thermal-state interval from the experimental pseudo-displacements retained after removal of spatially correlated assignment artefacts. To determine the identity-corrected 3D diffusivities and apparent activation energies, each of the 200 complete correction factor sets obtained from the simulations was applied separately to the corresponding experimental pseudo-diffusivities for Pd and Pt. Each application yielded 4 identity-corrected diffusivities for each species, which were used together for one Arrhenius fit.

For each species and correction factor set, the resulting 4 identity-corrected diffusivities were fitted to the Arrhenius relation $D = D_0 \exp\left(-\frac{E_a}{k_B T}\right)$, where $D_0$ is the pre-exponential factor, $E_a$ is the apparent activation energy, $k_B$ is the Boltzmann constant, and $T$ is the absolute temperature. Each thermal-state interval was represented by the annealing temperature of the later thermal state, corresponding to 400, 500, 600, and 650 °C.

To estimate the uncertainty in the temperature used for the Arrhenius fit, the measured and ideal thermal budgets were compared for each pulse–quench temperature–time profile (Supplementary Fig. 2). For each profile, the post-quench baseline temperature, $T_{\text{baseline}}$, was determined by averaging the measured temperatures during the unheated period. The measured thermal budget, $B_{\text{measured}}$, was calculated by integrating the temperature above this baseline over the recorded pulse–quench interval using the recorded data points. The ideal thermal budget was calculated using the target temperature of the corresponding pulse–quench heating cycle, $T_{\text{target}}$, and the temperature uncertainty was calculated from the difference between the measured and ideal thermal budgets as $B_{\text{ideal}} = (T_{\text{target}} - T_{\text{baseline}}) \times 900\ \text{s}$ and $\sigma_T = \frac{|B_{\text{measured}} - B_{\text{ideal}}|}{900\ \text{s}}$. This calculation was performed separately for the 4 temperature–time profiles. The resulting temperature uncertainties for 400, 500, 600 and 650 °C were 0.19, 0.30, 0.48 and 0.22 K, respectively.

For each species and correction factor set (indexed with $q$), the 4 identity-corrected diffusivities belonging to that set were fitted to the linearized Arrhenius relation between $\ln D$ and $1/T$ by weighted linear regression. Relative weights of $1/\sigma_{1/T}^2$ were used, where $\sigma_{1/T} = \sigma_T/T^2$. This regression first yielded the slope $m_q$ and intercept $b_q$. For each of the 4 temperature points, the fitted slope was then used to propagate the uncertainty in $1/T$ to $\ln D$ as $\sigma_{\ln D,q} = |m_q|\sigma_{1/T}$. The covariance matrix of the fitted intercept and slope was calculated as $(\mathbf{H}^{\text{T}}\mathbf{W}_q\mathbf{H})^{-1}$, where each row of $\mathbf{H}$ was $(1, 1/T)$, and the corresponding diagonal element of $\mathbf{W}_q$ was $1/\sigma_{\ln D,q}^2$. This covariance matrix was used directly without scaling by the residual mean squared error, thereby retaining only the contribution propagated from the assigned temperature uncertainties. The square roots of its diagonal elements were taken as the uncertainties $u_{b,q}$ and $u_{m,q}$ of the intercept and slope. The apparent activation energy and pre-exponential factor were calculated as $E_{\text{a},q} = -k_B m_q$ and $D_{0,q} = \exp(b_q)$, and their uncertainties were calculated by first-order uncertainty propagation as $u_{E_{\text{a}},q} = k_B u_{m,q}$ and $u_{D_0,q} = D_{0,q} u_{b,q}$.

For each thermal-state interval, the plotted value of $\ln D$ was the arithmetic mean of the values obtained from the 200 correction factor sets, and its vertical error bar was their sample standard deviation (Figs. 3a,b and 4a–d). The corresponding temperature uncertainties are negligible on the plotted scale and are therefore not visible. The reported value of $E_{\mathrm{a}}$ was the arithmetic mean of the values obtained from the 200 separate fits. The combined uncertainty arising from correction factor variation and temperature uncertainty was calculated as $u_{E_{\mathrm{a}}} = \sqrt{s_{E_a}^2 + \langle u_{E_a,q}^2 \rangle}$, where $s_{E_a}$ is the sample standard deviation of the 200 fitted values and $\langle u_{E_a,q}^2 \rangle$ is the arithmetic mean of the squared temperature-propagated uncertainties over the 200 correction factor sets. The uncertainties for $\ln(D_0)$ were similarly obtained. The plotted fitted line was calculated using the mean slope and mean intercept across the 200 fits. The complete fitting procedure was also repeated using the correction factor sets obtained from the spherical core–shell simulations for comparison (Supplementary Fig. 3j,k). The same fitting and uncertainty analysis was applied to the ground-truth simulation diffusivities, using the 20 trajectories for each particle geometry, with trajectory-to-trajectory variation replacing correction factor variation in the uncertainty calculation (Supplementary Fig. 3g,h).

**Transition matrices and nearest-layer transfer**

To quantify the redistribution of atoms between atomic layers, source-normalized transition matrices were constructed separately for each species and each of the 4 thermal-state intervals. For each pseudo-trajectory, the layer index in the earlier thermal state was defined as the source layer $l$, and that in the later thermal state was defined as the destination layer $k$. Let $N_{kl}$ denote the number of pseudo-trajectories from $l$ to $k$ and let $n_l = \sum_{k'} N_{k'l}$, summed over all destination layers $k'$. The transition probability was calculated as $M_{kl} = \frac{N_{kl}}{n_l}$, and the values $M_{kl}$ formed the source-normalized transition matrix. For source layers containing no retained pseudo-trajectories, the corresponding matrix columns were left undefined.

The unnormalized nearest-layer transfer across the boundary between the layers $l$ and $l + 1$ was calculated as $N_{l+1,l} - N_{l,l+1}$, and the corresponding normalized nearest-layer transfer was calculated as $\frac{N_{l+1,l} - N_{l,l+1}}{n_l + n_{l+1}}$. Positive and negative values represented net outward and inward transfer, respectively.

The complete transition matrices were calculated over $l = -9$ to 9. For visualization only, a layer was omitted from the displayed range if it contained less than 3% of the source-layer atoms in the retained pseudo-trajectories for every thermal-state interval in both species. This avoided extending the plots to consistently sparsely populated layers, where normalization over a small source population could overemphasize individual pseudo-trajectories. This criterion yielded the displayed range $l = -6$ to 5 (Supplementary Fig. 4a). This range was applied to the source-normalized transition matrices, transition probability plots for the principal matrix diagonals (Supplementary Fig. 4j,k) and nearest-layer transfer plots.

**Facet-resolved analysis**

To compare the 3D mobility and layer redistribution associated with the {100} and {111} facets, separate sets of retained pseudo-trajectories were constructed for the {100} and {111} facet families using the facet assignments defined above. A pseudo-trajectory was included in a facet family set only when its source and destination endpoints were assigned to the same resolved individual facet belonging to that family. Pseudo-trajectories whose exact facet assignment was unresolved or whose endpoints belonged to different individual facets, including different facets of the same family, were not included. For each facet family set, pseudo-diffusivities were calculated from the selected pseudo-trajectories and corrected using the correction factors obtained from the simulations. The Arrhenius fitting, uncertainty analysis, construction of transition matrices and nearest-layer transfer calculations were performed as described above.

To account for the different numbers of directed nearest-neighbour connections between adjacent layers in the ideal fcc geometry, an additional correction was applied to the facet-resolved nearest-layer transfers after source-layer normalization. For transfer across a specified adjacent-layer boundary, 4 of the 12 nearest-neighbour directions connect adjacent {100} layers, whereas 3 of the 12 directions connect adjacent {111} layers. The normalized transfers were therefore divided by 1/3 for the {100} facets and by 1/4 for the {111} facets, respectively.

**Use of large language model**

The authors used ChatGPT to assist with manuscript drafting and language revision and with the development of data-analysis and plotting scripts. The authors critically reviewed and verified the resulting text, code and analyses and take full responsibility for the work.

# References

52. Kwon, Y. *et al.* CO adsorption-induced deposition: A facile and precise synthesis route for core–shell catalysts. *ACS Nano* **19**, 39520–39530 (2025).
53. Meyer, J. C. *et al.* Accurate Measurement of Electron Beam Induced Displacement Cross Sections for Single-Layer Graphene. *Phys. Rev. Lett.* **108**, 196102 (2012).
54. Jimenez, C. M., Lowe, L. F., Burke, E. A. & Sherman, C. H. Radiation Damage in Pd Produced by 1-3-MeV Electrons. *Phys. Rev.* **153**, 735–740 (1967).
55. Jung, P., Chaplin, R. L., Fenzl, H. J., Reichelt, K. & Wombacher, P. Anisotropy of the Threshold Energy for Production of Frenkel Pairs in Copper and Platinum. *Phys. Rev. B* **8**, 553–561 (1973).
56. Lee, J., Jeong, C., Lee, T., Ryu, S. & Yang, Y. Direct Observation of Three-Dimensional Atomic Structure of Twinned Metallic Nanoparticles and Their Catalytic Properties. *Nano Lett.* **22**, 665–672 (2022).
57. Jeong, C. *et al.* Revealing the three-dimensional arrangement of polar topology in nanoparticles. *Nat. Commun.* **15**, 3887 (2024).
58. Lee, J., Jeong, C. & Yang, Y. Single-atom level determination of 3-dimensional surface atomic structure via neural network-assisted atomic electron tomography. *Nat. Commun.* **12**, 1962 (2021).
59. Hong, J. *et al.* Metastable hexagonal close-packed palladium hydride in liquid cell TEM. *Nature* **603**, 631–636 (2022).

60. Jeong, C. *et al.* Atomic-scale 3D structural dynamics and functional degradation of Pt alloy nanocatalysts during the oxygen reduction reaction. *Nat. Commun.* **16**, 8026 (2025).
61. Lee, E. *et al.* Suppressing Metal Dissolution in Multi-Grained Catalysts Through Intragrain Atomic Ordering for Stable Fuel Cells. *Adv. Mater.* **37**, 2504059 (2025).
62. Dabov, K., Foi, A., Katkovnik, V. & Egiazarian, K. Image Denoising by Sparse 3-D Transform-Domain Collaborative Filtering. *IEEE Trans. Image Process.* **16**, 2080–2095 (2007).
63. Scott, M. C. *et al.* Electron tomography at 2.4-ångström resolution. *Nature* **483**, 444–447 (2012).
64. Frank, J. & McEwen, B. F. Alignment by Cross-Correlation. in *Electron Tomography: Three-Dimensional Imaging with the Transmission Electron Microscope* (ed. Frank, J.) 205–213 (Plenum Press, New York, 1992).
65. Lewis, J. P. *Fast Normalized Cross-Correlation*. https://scribblethink.org/Work/nvisionInterface/nip.pdf (1995).
66. Pham, M., Yuan, Y., Rana, A., Osher, S. & Miao, J. Accurate real space iterative reconstruction (RESIRE) algorithm for tomography. *Sci. Rep.* **13**, 5624 (2023).
67. Tian, X. *et al.* Correlating the three-dimensional atomic defects and electronic properties of two-dimensional transition metal dichalcogenides. *Nat. Mater.* **19**, 867–873 (2020).
68. Yang, Y. *et al.* Atomic-scale identification of active sites of oxygen reduction nanocatalysts. *Nat. Catal.* **7**, 796–806 (2024).
69. Edelsbrunner, H. & Mücke, E. P. Three-dimensional alpha shapes. *ACM Trans. Graph. TOG* **13**, 43–72 (1994).
70. Gurobi Optimization, LLC. Gurobi Optimizer Reference Manual. (2026).
71. Ophus, C. A fast image simulation algorithm for scanning transmission electron microscopy. *Adv. Struct. Chem. Imaging* **3**, 13 (2017).
72. Pryor, A., Ophus, C. & Miao, J. A streaming multi-GPU implementation of image simulation algorithms for scanning transmission electron microscopy. *Adv. Struct. Chem. Imaging* **3**, 15 (2017).
73. Kirkland, E. J. *Advanced Computing in Electron Microscopy*. (Springer, New York, 2010).
74. Rangel DaCosta, L. *et al.* Prismatic 2.0 – Simulation software for scanning and high resolution transmission electron microscopy (STEM and HRTEM). *Micron* **151**, 103141 (2021).
75. Meyer, M., Desbrun, M., Schröder, P. & Barr, A. H. Discrete Differential-Geometry Operators for Triangulated 2-Manifolds. in *Visualization and Mathematics III* (eds Hege, H.-C. & Polthier, K.) 35–57 (Springer, Berlin, Heidelberg, 2003). doi:10.1007/978-3-662-05105-4_2.

## Acknowledgements

The authors thank Dr. Eva Bladt for her assistance in securing MEMS chips and Prof. Jae-Hyung Jeon for valuable discussions. The electron microscopy measurements were conducted using electron microscopes (FEI Titan G2 and ThermoFisher Spectra Ultra) in the KAIST Analysis Center for Research Advancement (KARA), which was partially supported by KAIST singularity professor program. Excellent support by Hyung Bin Bae, Jin-Seok Choi and the staff of KARA is gratefully acknowledged.

## Author contributions

Y.Y. conceived the idea and directed the study. Y.K., Y.W., G.-G.P., E.L. and S.W.H. synthesized the Pd@Pt core–shell nanoparticles. H.J., S.H., J.O., D.P. and Y.Y. designed and performed the pulse-quench electron microscopy experiments. D.P., H.J., and Y.Y. conducted the tomography data analysis. D.P. and Y.Y. conducted lattice-exchange simulation and atom-matching analysis. D.P., H.J., S.H., Y.K. and Y.Y. wrote the manuscript. All authors commented on the manuscript.

## Funding statement

This research was mainly supported by Samsung Science and Technology Foundation (SSTF-BA2201-05). S.W.H. acknowledges the support from the National Research Foundation of Korea (NRF) grant funded by the Korean government (MSIT) (RS-2024-00350471 and RS-2026-25582850).

# Supplementary Information

for

Three-dimensional atom-by-atom measurement of interface diffusion dynamics

Doojin Park[1,†], Hyesung Jo[1,†], Yongmin Kwon[2,3], Seokjo Hong[1], Jaewhan Oh[1], Youngjoo Whang[2,3], Eunjik Lee[3,4,5], Gu-Gon Park[3,4,5], Sang Woo Han[2,*] and Yongsoo Yang[1,6,**]

[1] *Department of Physics, Korea Advanced Institute of Science and Technology (KAIST), Daejeon 34141, Republic of Korea*

[2] *Department of Chemistry, Korea Advanced Institute of Science and Technology (KAIST), Daejeon 34141, Republic of Korea*

[3] *Hydrogen Fuel Cell Laboratory, Korea Institute of Energy Research (KIER), Daejeon 34129, Republic of Korea*

[4] *Department of Energy Engineering, University of Science and Technology (UST), Daejeon 34113, Republic of Korea*

[5] *Graduate School of Energy Science and Technology (GEST), Chungnam National University, Daejeon 34134, Republic of Korea*

[6] *Graduate School of Semiconductor Technology, School of Electrical Engineering, Korea Advanced Institute of Science and Technology (KAIST), Daejeon 34141, Republic of Korea*

[†] These authors contributed equally to this work.
Corresponding author email: *sangwoohan@kaist.ac.kr, **yongsoo.yang@kaist.ac.kr

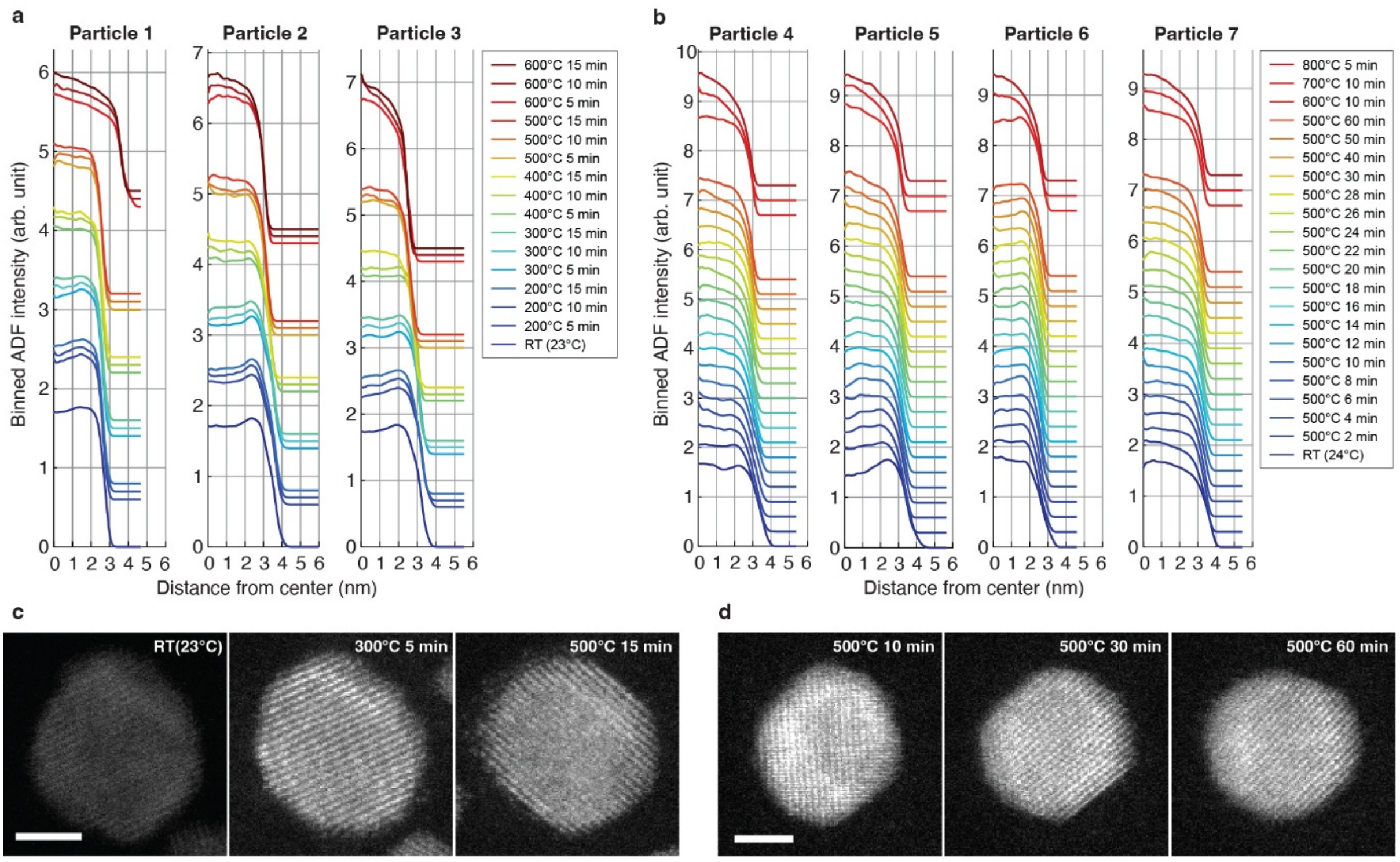

**Supplementary Figure 1 | Pulse–quench ADF-STEM screening of Pd@Pt nanoparticle interdiffusion. a**, Temperature-dependent pulse–quench ADF-STEM screening of three Pd@Pt nanoparticles. Each particle was imaged at room temperature and then subjected to repeated heat–hold–quench–image cycles at 200, 300, 400, 500 and 600 °C, with accumulated dwell times of 5, 10 and 15 min at each temperature. Binned radial ADF intensity profiles are plotted as a function of distance from the nanoparticle centre. The initially clear core–shell contrast, characterized by a lower-intensity Pd-rich core and higher-intensity Pt-rich shell, progressively diminishes with increasing temperature and dwell time. **b**, Time-dependent pulse–quench ADF-STEM screening of four Pd@Pt nanoparticles at 500 °C for accumulated dwell times up to 60 min, followed by additional heating at 600, 700 and 800 °C. The radial intensity profiles show gradual loss of the core–shell contrast at 500 °C and accelerated intermixing at higher temperatures. **c**,**d**, Representative ADF-STEM images from the temperature-dependent (c) and time-dependent (d) screening experiments. The images show that the Pt-shell contrast decreases as interfacial diffusion proceeds, while the overall nanoparticle morphology remains largely preserved under the conditions selected for tomography. These pulse–quench screening experiments were used to identify the final AET annealing sequence of 900 s at 300, 400, 500, 600 and 650 °C, which activates interfacial diffusion while limiting global shape changes. Scale bars, 2 nm.

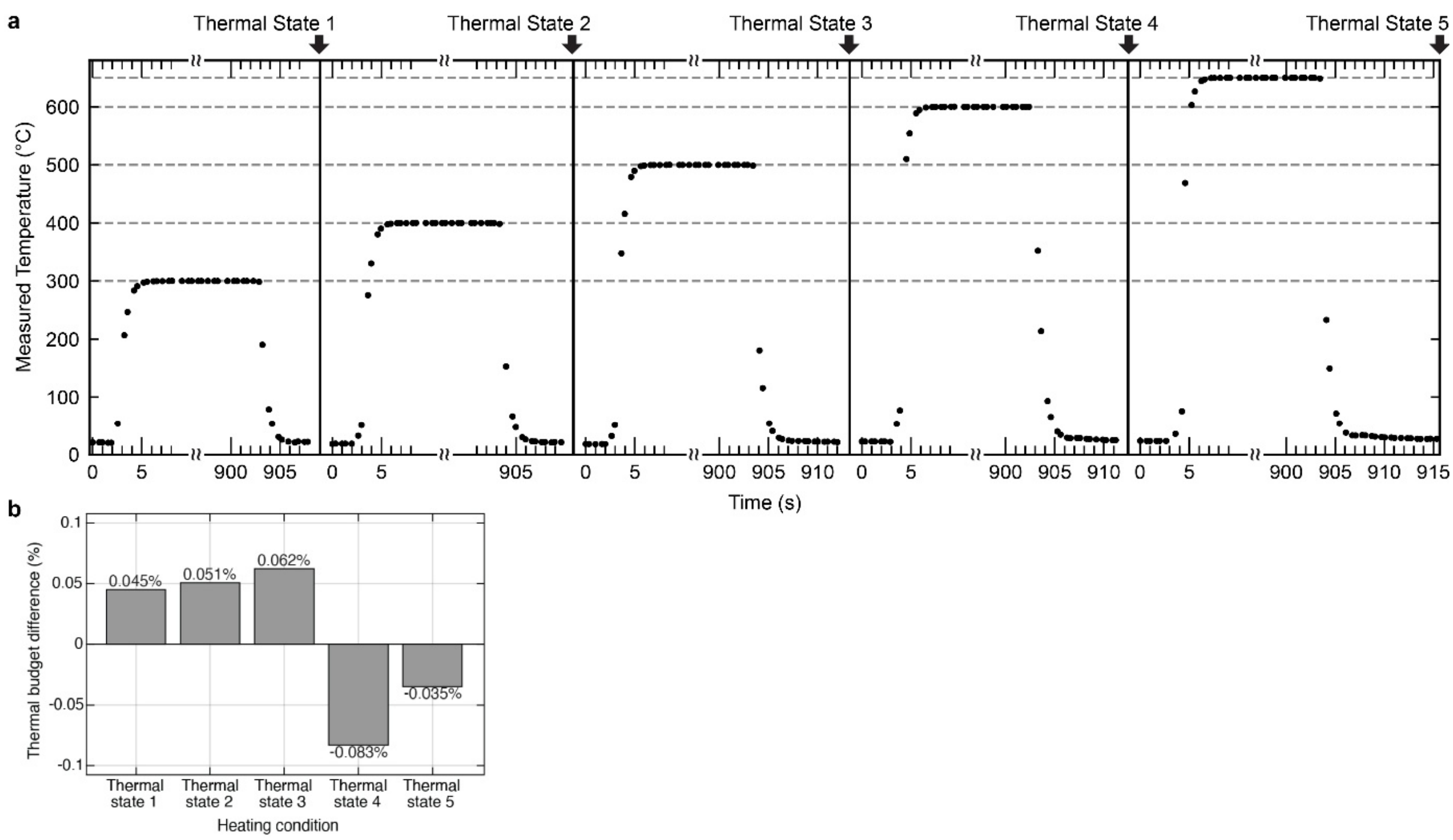

**Supplementary Figure 2 | Fast temperature response of the MEMS heating chip during experiments. a**, Black dots show the measured temperature of the MEMS chip during sequential heating cycles to 300, 400, 500, 600, and 650 °C, with return to room temperature after each cycle. Grey dashed horizontal lines indicate the target temperatures. The time axis is broken to omit long dwell periods while preserving the local time scale around each heating and cooling ramp; tick labels show elapsed time in seconds within each temperature cycle. The sample reaches each target temperature and cools back to near room temperature within a few seconds, demonstrating that the diffusion experiments are not dominated by slow thermal ramping. Arrows mark the five thermal states used for structural analysis. **b**, Thermal-budget accuracy of MEMS pulse–quench heating. Relative thermal-budget difference for each pulse–quench thermal state, calculated by integrating the measured sample temperature over the full heating window of each pulse–quench sequence and comparing it with the ideal isothermal budget, $(T_{\text{target}} - T_{\text{baseline}}) \times 900$ s. Positive and negative values indicate slight over- and under-delivery compared to the nominal temperature–time budget, respectively. For all five thermal states, the deviation remains below 0.1%, confirming that the MEMS pulse–quench protocol provides well-defined thermal exposures for quantitative diffusion analysis.

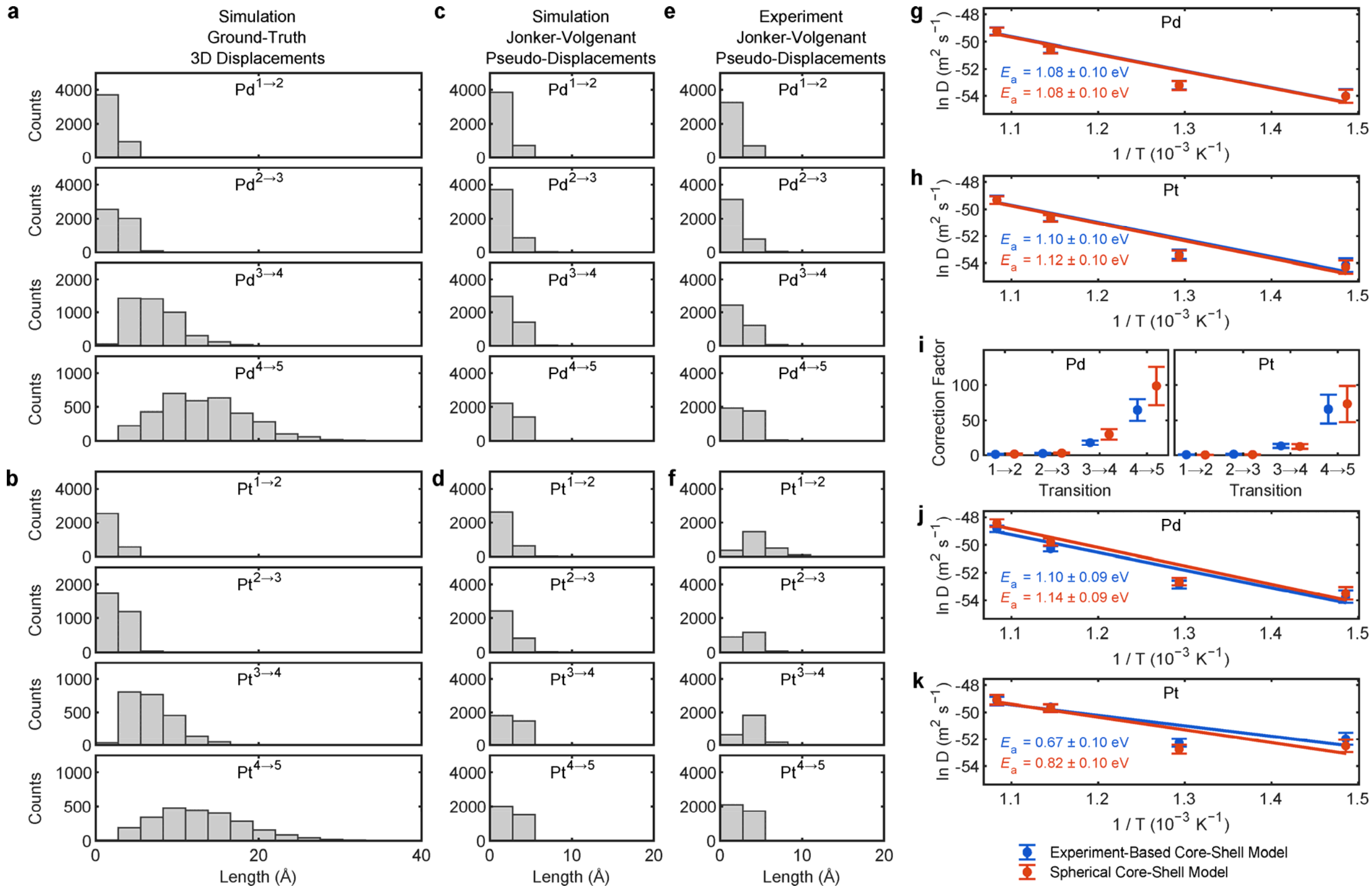

**Supplementary Figure 3 | Robustness of the simulation calibration to core–shell geometry. a**,**b**, Distributions of displacement magnitudes for Pd (a) and Pt (b) obtained from the labelled ground-truth 3D displacements in the lattice-exchange simulations over the four consecutive thermal-state intervals. **c**,**d**, Corresponding Jonker–Volgenant pseudo-displacement distributions after removal of atom identities. **e**,**f**, Experimental Jonker–Volgenant pseudo-displacement distributions. **g**,**h**, Arrhenius plots of the labelled simulation diffusivities for Pd (g) and Pt (h), comparing the experiment-based core–shell model used in the main analysis (blue) with an ideal spherical core–shell model containing similar numbers of Pd and Pt atoms (red). **i**, Interval-specific correction factors for Pd and Pt obtained from the two core–shell models. Error bars denote standard deviations across the simulation ensembles. **j**,**k**, Experimental Pd (j) and Pt (k) diffusivities corrected using the corresponding simulation ensembles; the blue results are those reported in the main text. Solid lines show the Arrhenius relation, with the corresponding activation energies indicated in matching colours. Comparison with the labelled ground-truth displacement distributions shows that the pseudo-trajectories systematically compress the displacement distributions, while the simulated pseudo-displacement distributions are broadly consistent with the experimental ones. The similar correction factors and activation energies show that the calibration is robust to the detailed core–shell morphology and geometry.

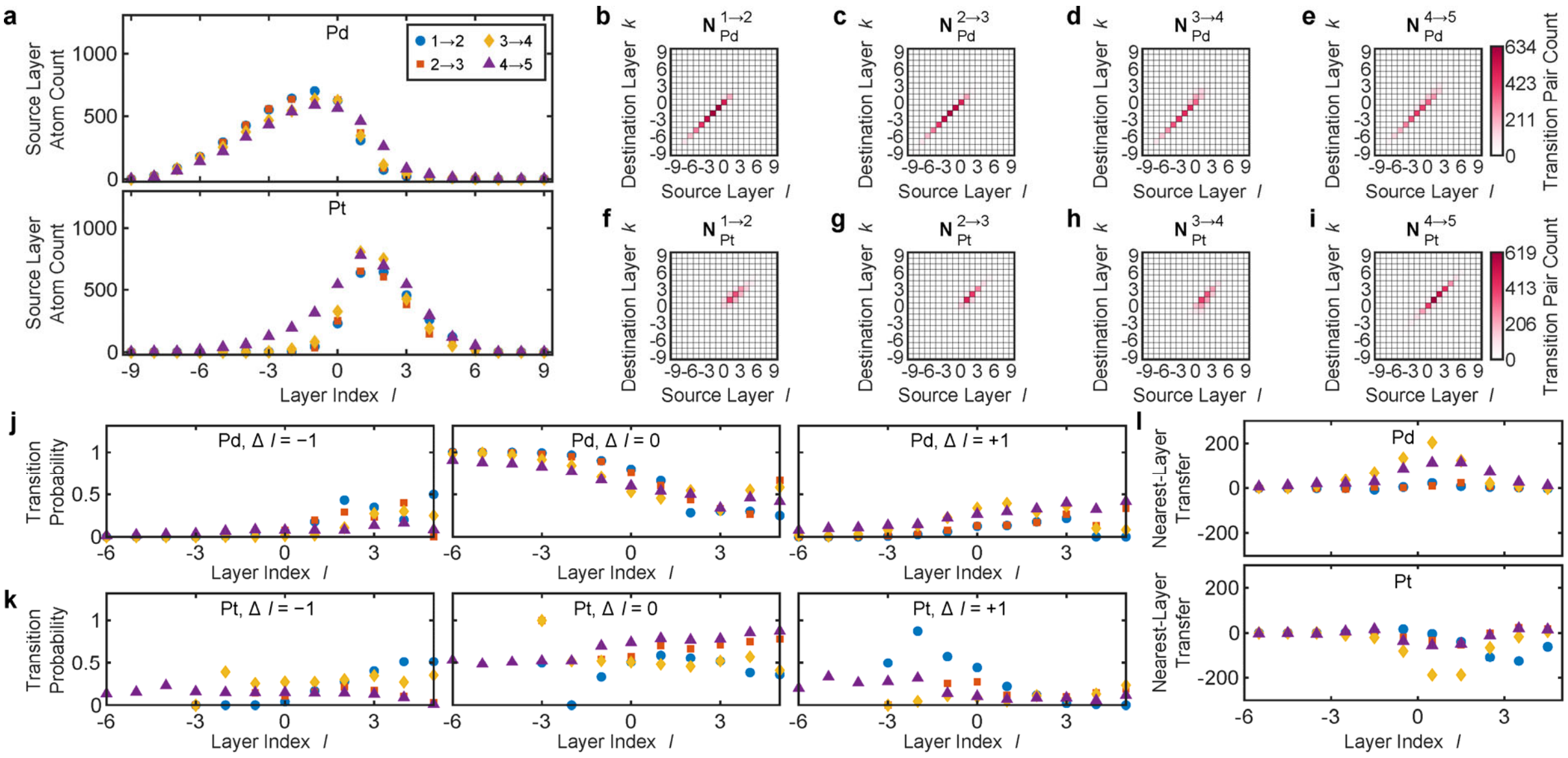

**Supplementary Figure 4 | Layer-transition counts and nearest-layer components. a**, Numbers of Pd and Pt atoms retained in the final Jonker–Volgenant assignment for each source layer and thermal-state interval, showing the Pd-rich core ($l < 0$) and Pt-rich shell ($l > 0$). **b**–**e**, Raw Pd transition-pair counts from source layer $l$ to destination layer $k$ over consecutive thermal-state intervals. **f**–**i**, Corresponding Pt transition-count matrices. Source-normalized transition matrices and interlayer transfers in Fig. 3 can all be obtained via proper normalization of these raw count matrices (Methods). **j**, Pd transition probabilities for the three principal matrix diagonals, $\Delta l = k - l = -1$, 0 and +1, respectively. Core Pd predominantly remains in the same layer, whereas neighbouring-layer transitions are concentrated near and outside the interface, with outward transitions ($\Delta l = +1$) generally exceeding inward transitions ($\Delta l = -1$). **k**, Corresponding Pt probabilities. Pt shows a broader shell-side distribution, reflecting additional shell reshaping and near-surface motion, while inward transitions tend to dominate near the interface. Missing points in the $l < -3$ region for some thermal-state intervals correspond to unpopulated source layers. **l**, Unnormalized nearest-layer transfer for Pd and Pt, calculated from the difference between outward and inward adjacent-layer transition counts. Positive and negative values denote net outward and inward transfer, respectively. The raw transfers are concentrated near the interface, because the deep Pd core and outer Pt shell yield predominantly same-layer Jonker–Volgenant assignments, whereas cross-layer matched pairs are most frequent in the interfacial region.

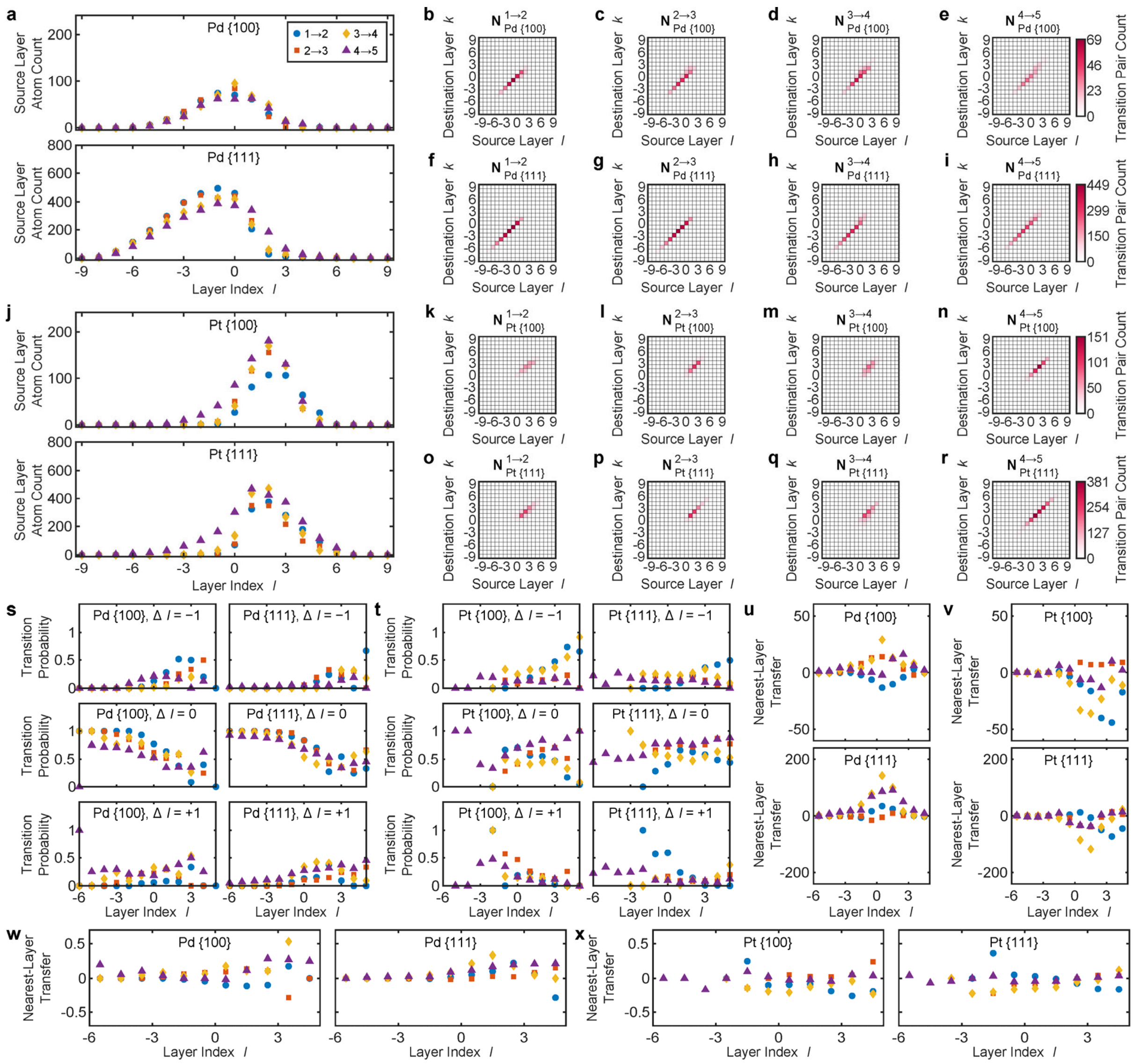

**Supplementary Figure 5 | Facet-resolved layer-transition counts and nearest-layer components. a**, Numbers of Pd atoms retained in the final Jonker–Volgenant assignment for each source layer and thermal-state interval, separated into {100}- and {111}-associated regions. **b**–**e**, Raw Pd transition-pair counts from source layer $l$ to destination layer $k$ for {100} regions over consecutive thermal-state intervals. **f**–**i**, Corresponding Pd counts for {111} regions. **j**, Source-layer counts for Pt in the two facet families. **k**–**n**, Raw Pt transition-pair counts for {100} regions. **o**–**r**, Corresponding Pt counts for {111} regions. Source-normalized transition matrices and interlayer transfers in Fig. 4 can all be obtained via proper normalization of these raw count matrices (Methods). **s**, Pd transition probabilities along the three principal matrix diagonals, $\Delta l = k - l = -1$, 0 and +1, for both facet families. Core Pd predominantly remains in the same layer, whereas neighbouring-layer transitions are concentrated near and outside the interface, with outward transitions generally exceeding inward transitions. The smaller cross-layer contribution in {111} regions is consistent with their six cross-layer nearest-neighbour pathways, compared with eight for {100}. **t**, Corresponding Pt probabilities, showing a less systematic tendency owing to additional shell reshaping and near-surface motion. **u**,**v**, Unnormalized nearest-layer transfer for Pd (u) and Pt (v). Positive and negative values denote net outward and inward transfer, respectively; the raw transfers are concentrated near the interface, where cross-layer matched pairs are most frequent. **w**,**x**, Nearest-layer transfer for Pd (w) and Pt (x), normalized by the number of atoms available for adjacent-layer transfer but not by the ideal fcc cross-layer connectivity (Methods). Before this connectivity normalization, {100} regions show greater interlayer transfer than {111} regions, most clearly for Pd between thermal states 4 and 5, consistent with their larger number of cross-layer pathways.

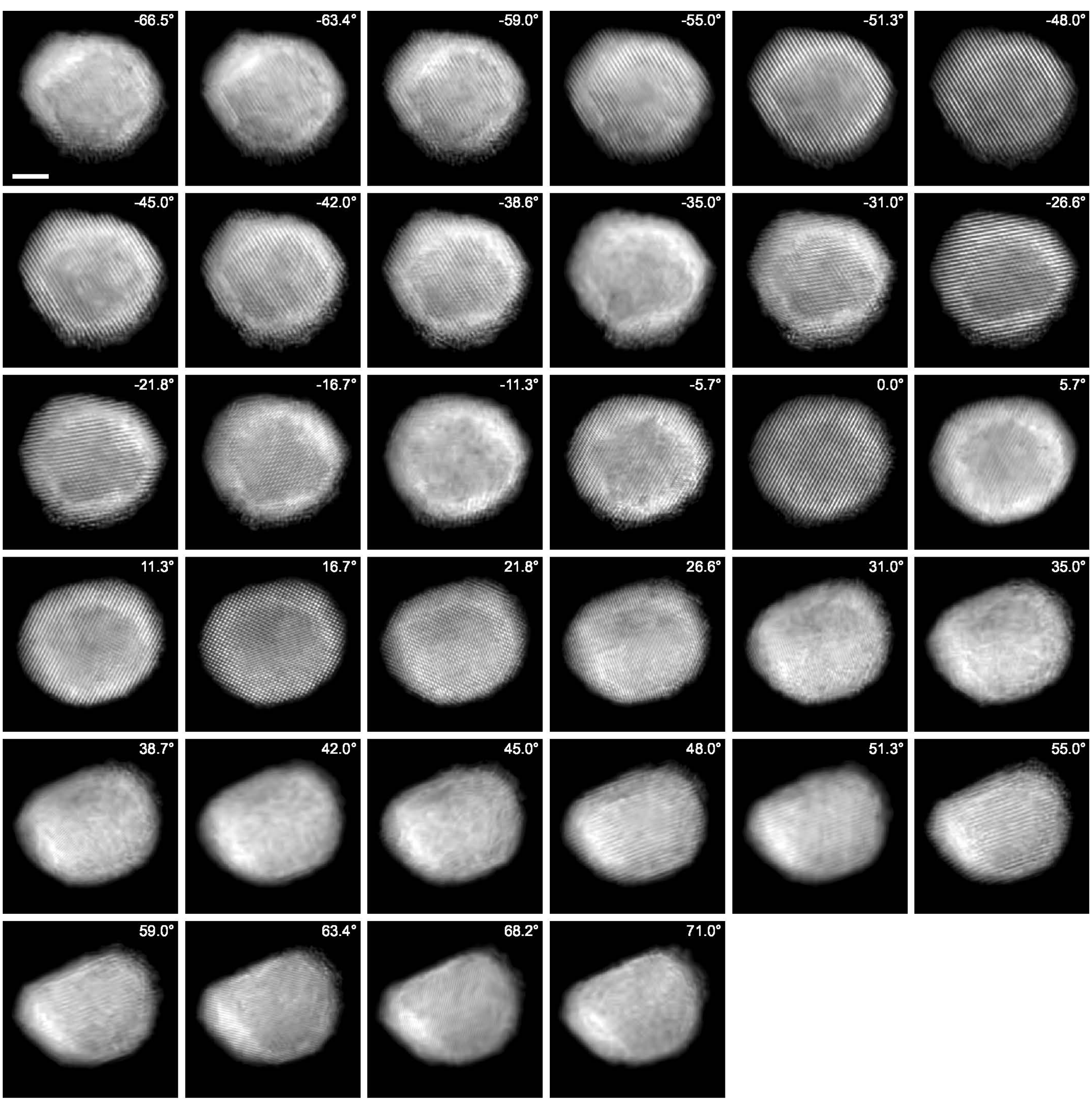

**Supplementary Figure 6 | An experimental tomographic tilt series of the Pd@Pt nanoparticle after MEMS-chip-based rapid thermal pulsing at 300 °C (thermal state 1).** A total of 34 tilt series images were acquired from an ADF-STEM experiment and post-processed as described in the Methods. The corresponding tilt angle for each projection is shown at the top right corner of each image. Scale bar, 2 nm.

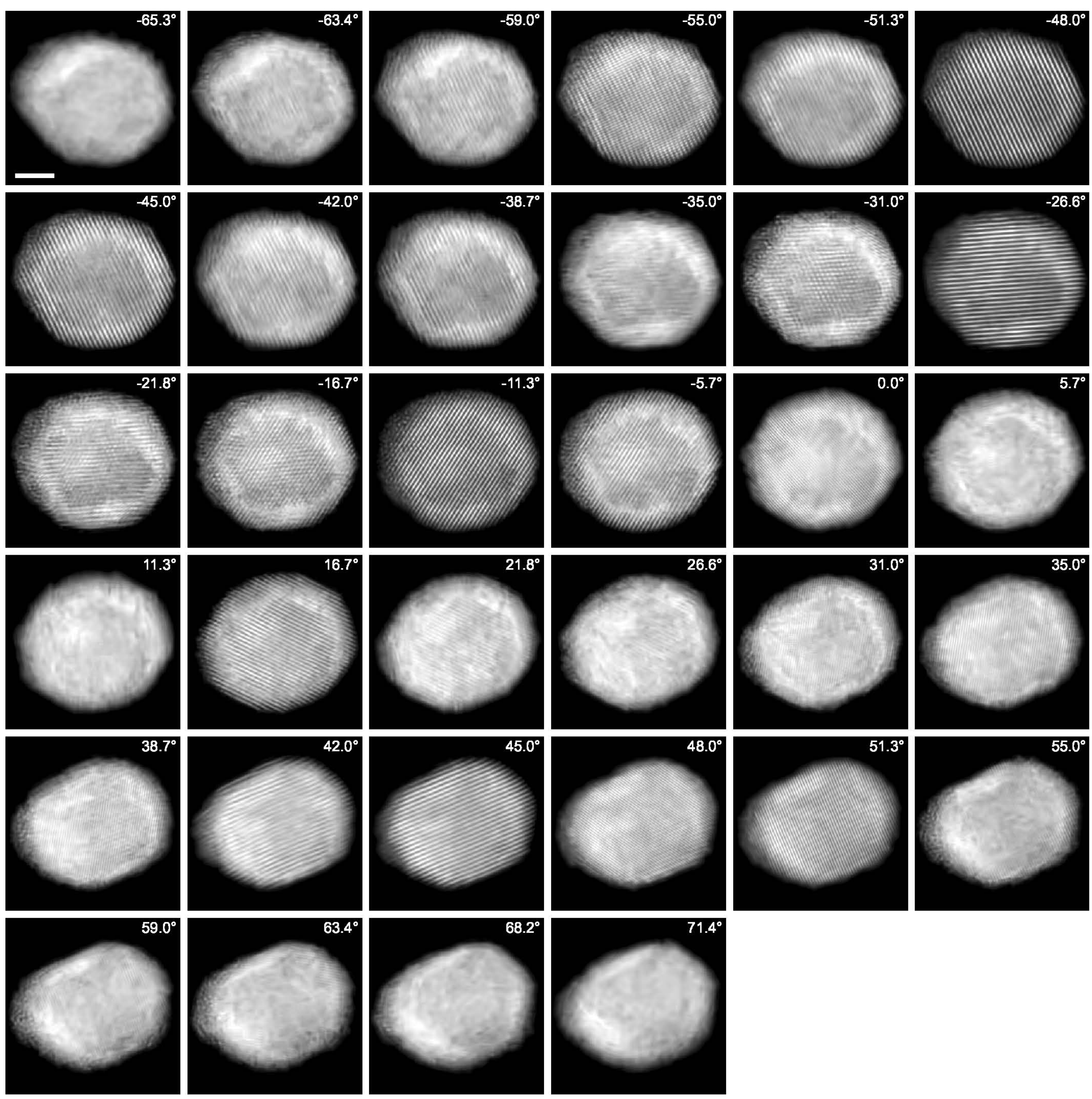

**Supplementary Figure 7 | An experimental tomographic tilt series of the Pd@Pt nanoparticle after MEMS-chip-based rapid thermal pulsing at 400 °C (thermal state 2).** A total of 34 tilt series images were acquired from an ADF-STEM experiment and post-processed as described in the Methods. The corresponding tilt angle for each projection is shown at the top right corner of each image. Scale bar, 2 nm.

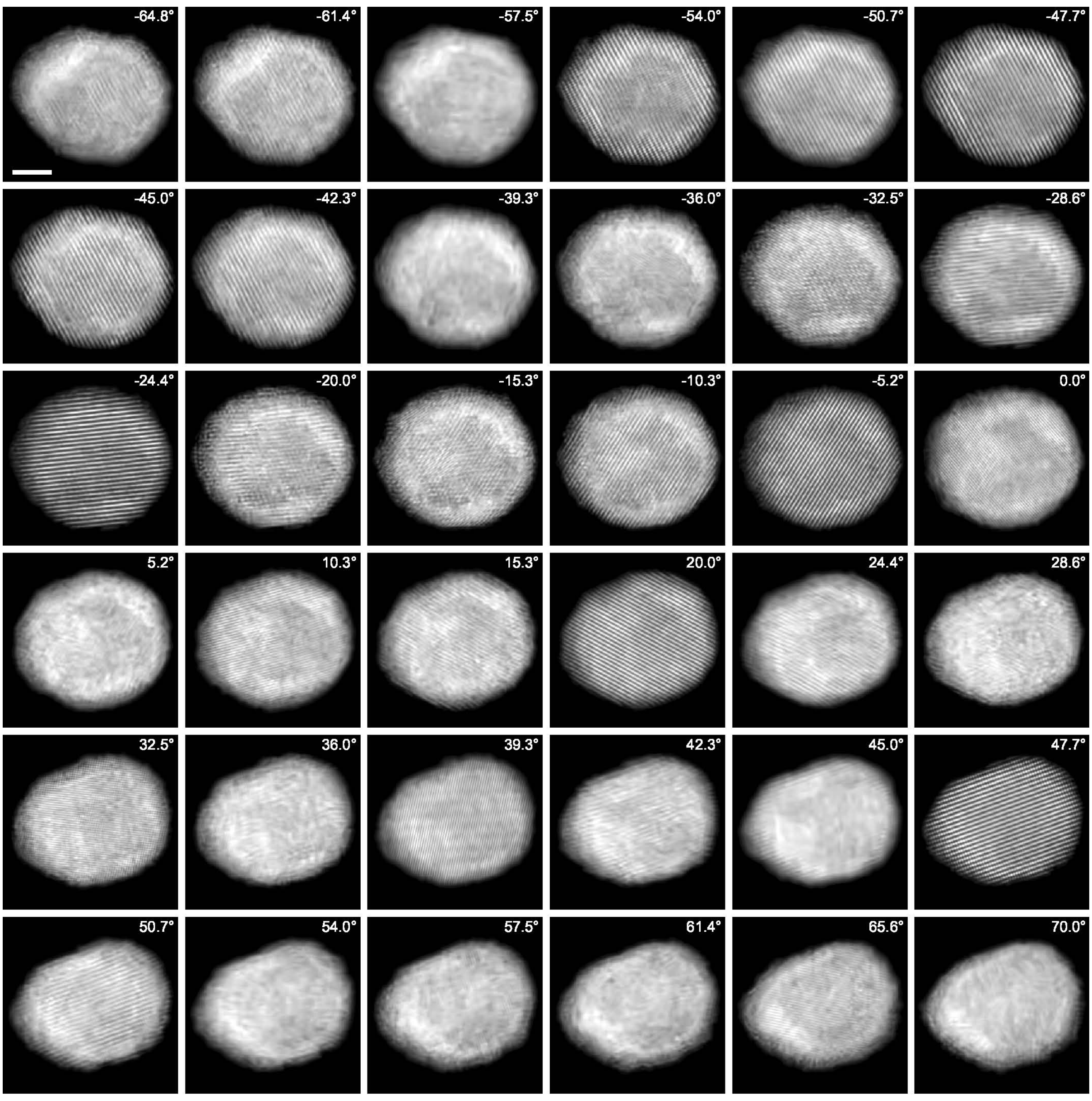

**Supplementary Figure 8 | An experimental tomographic tilt series of the Pd@Pt nanoparticle after MEMS-chip-based rapid thermal pulsing at 500 °C (thermal state 3).** A total of 36 tilt series images were acquired from an ADF-STEM experiment and post-processed as described in the Methods. The corresponding tilt angle for each projection is shown at the top right corner of each image. Scale bar, 2 nm.

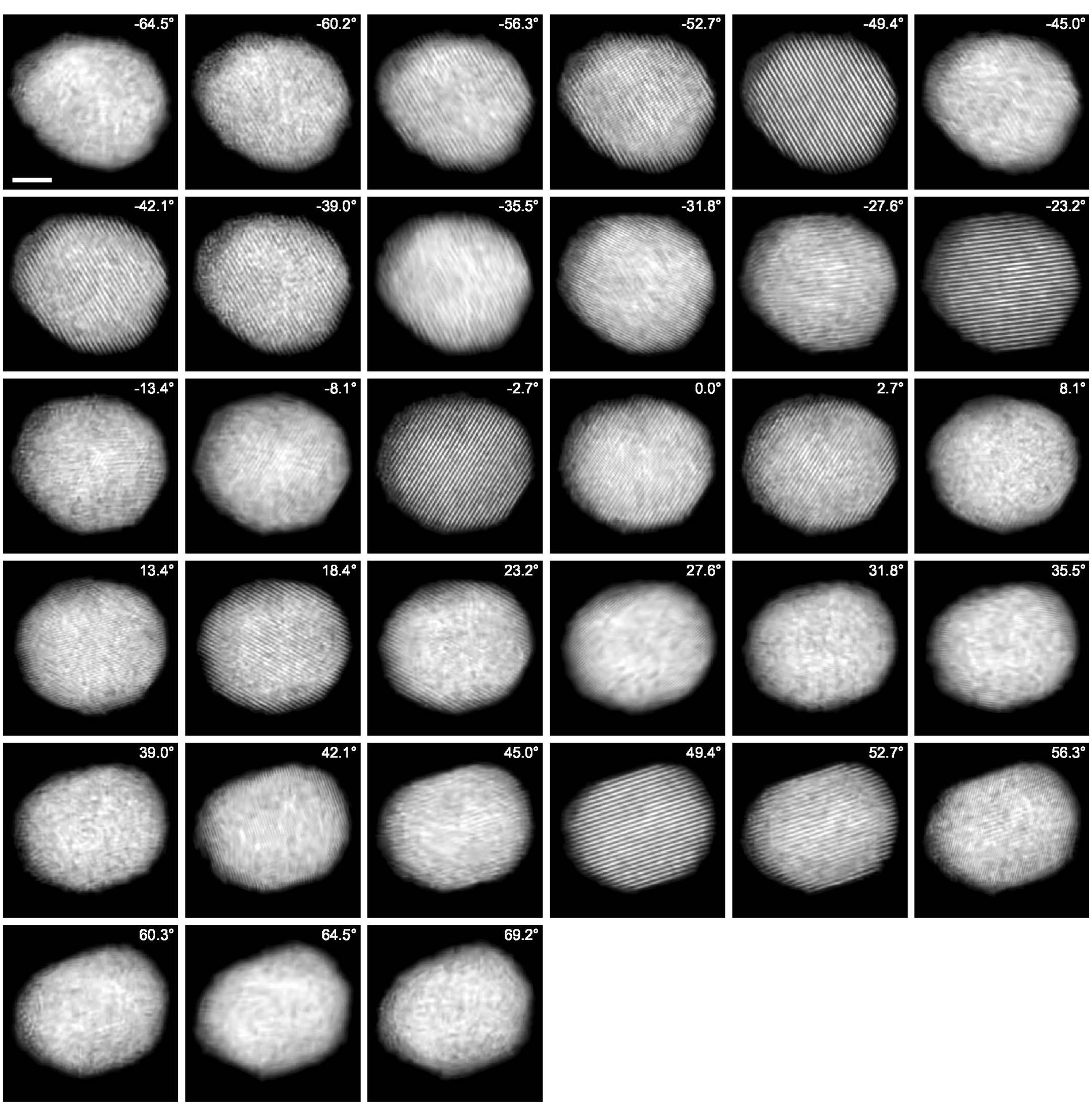

**Supplementary Figure 9 | An experimental tomographic tilt series of the Pd@Pt nanoparticle after MEMS-chip-based rapid thermal pulsing at 600 °C (thermal state 4).** A total of 33 tilt series images were acquired from an ADF-STEM experiment and post-processed as described in the Methods. The corresponding tilt angle for each projection is shown at the top right corner of each image. Scale bar, 2 nm.

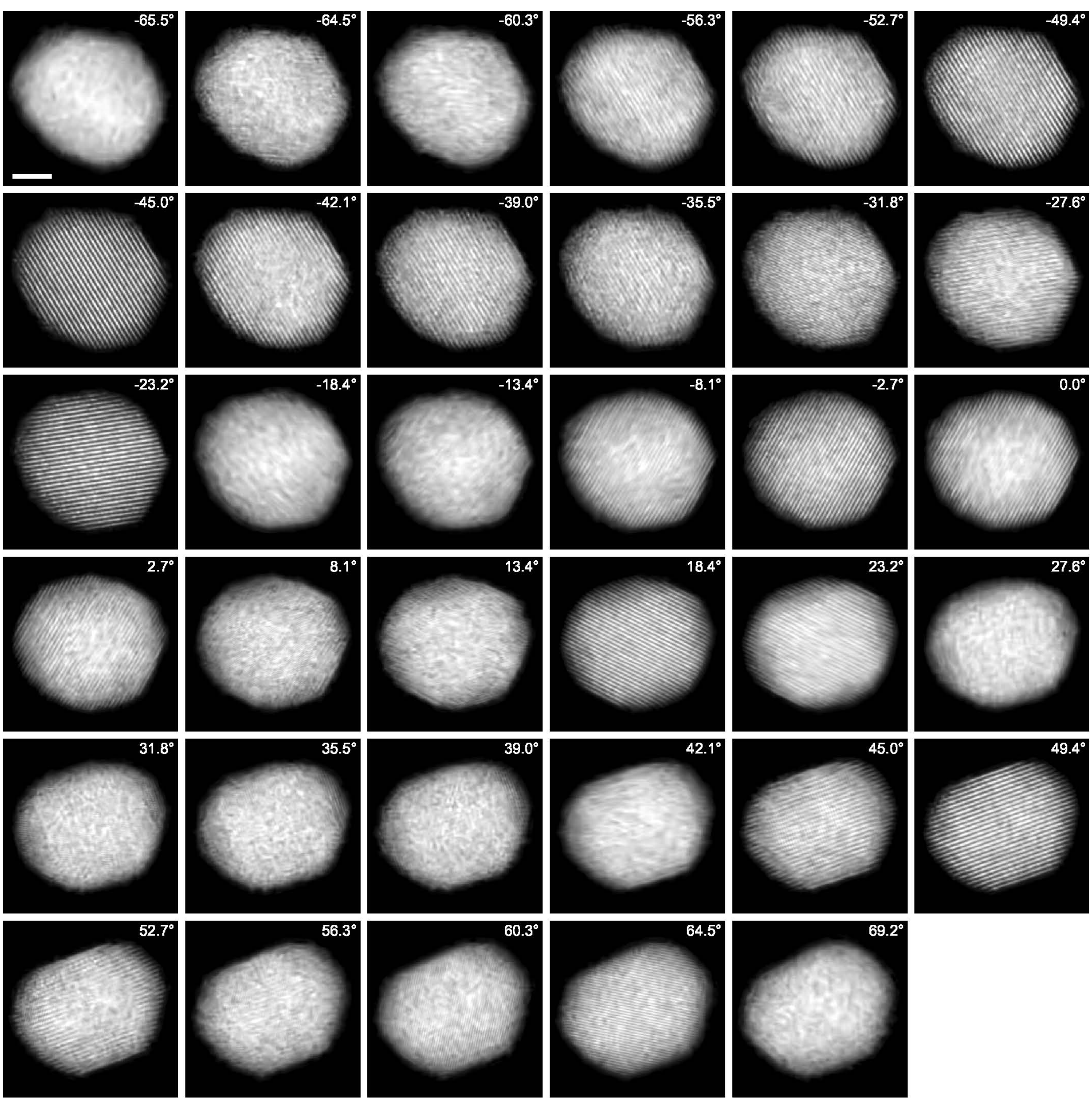

**Supplementary Figure 10 | An experimental tomographic tilt series of the Pd@Pt nanoparticle after MEMS-chip-based rapid thermal pulsing at 650 °C (thermal state 5).** A total of 35 tilt series images were acquired from an ADF-STEM experiment and post-processed as described in the Methods. The corresponding tilt angle for each projection is shown at the top right corner of each image. Scale bar, 2 nm.

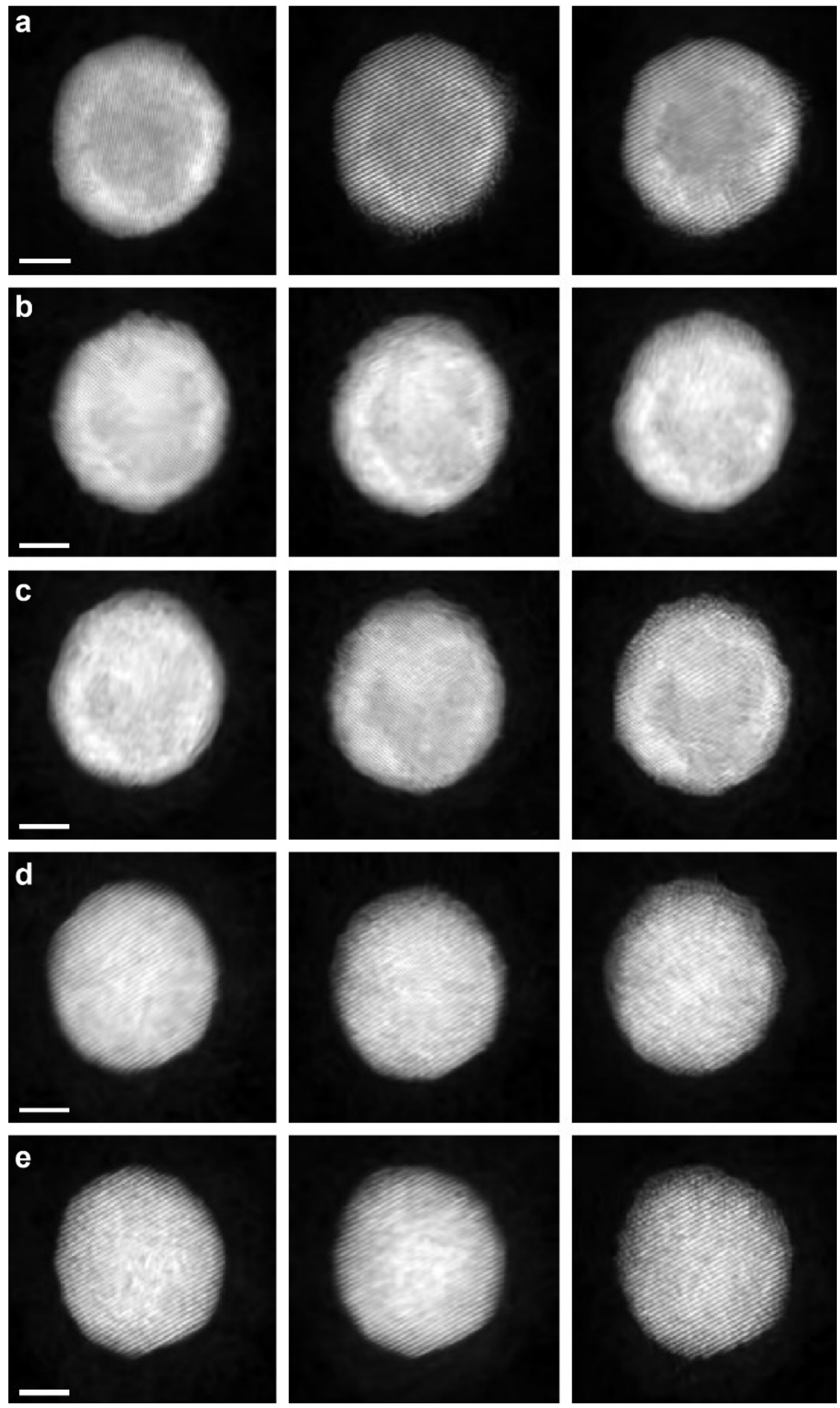

**Supplementary Figure 11 | Experimental zero-degree projections acquired at the beginning (left column), in the middle (middle column), and at the end (right column) of the tilt-series acquisition. a–e**, Zero-degree images of the tilt-series measured after rapid thermal pulsing at 300 °C (**a**), 400 °C (**b**), 500 °C (**c**), 600 °C (**d**), and 650 °C (**e**). Scale bars, 2 nm.

**Supplementary Table 1 | Experimental conditions, analysis parameters and tracing results of the Pd@Pt nanoparticle at different thermal states.**

| | Thermal state 1 (after 300 °C) | Thermal state 2 (after 400 °C) | Thermal state 3 (after 500 °C) | Thermal state 4 (after 600 °C) | Thermal state 5 (after 650 °C) |
|---|---|---|---|---|---|
| **STEM data acquisition** | | | | | |
| Electron microscope type | FEI Titan | TFS Spectra Ultra | TFS Spectra Ultra | TFS Spectra Ultra | TFS Spectra Ultra |
| Acceleration voltage (kV) | 300 | 300 | 300 | 300 | 300 |
| Convergence semi-angle (mrad) | 18.0 | 21.4 | 21.4 | 21.4 | 21.4 |
| Detector inner angle (mrad) | 39 | 39 | 39 | 39 | 39 |
| Detector outer angle (mrad) | 200 | 200 | 200 | 200 | 200 |
| Pixel size (Å) | 0.362 | 0.331 | 0.331 | 0.331 | 0.331 |
| Beam current (pA) | 15 | 15 | 15 | 15 | 15 |
| Dwell time (μs) | 3 | 3 | 3 | 3 | 3 |
| # of consecutive images | 3 | 3 | 3 | 3 | 3 |
| # of projections | 34 | 34 | 36 | 33 | 35 |
| Tilt angle range (°) | −66.5<br>+71.0 | −65.3<br>+71.4 | −64.8<br>+70.0 | −64.5<br>+69.2 | −65.5<br>+69.2 |
| Electron dose ($10^5$ $e$/Å$^2$) | 2.2 | 2.6 | 2.8 | 2.5 | 2.7 |
| **3D reconstruction** | | | | | |
| Algorithm | RESIRE | RESIRE | RESIRE | RESIRE | RESIRE |
| # of iterations | 1000 | 1000 | 1000 | 1000 | 1000 |
| Oversampling ratio | 4 | 4 | 4 | 4 | 4 |
| **Tracing result** | | | | | |
| # of atoms | | | | | |
| Pd | 6414 | 6414 | 6414 | 6414 | 6414 |
| Pt | 6381 | 6381 | 6381 | 6381 | 6381 |
| B factor (Å$^2$) | | | | | |
| Pd | 9.1 | 8.3 | 7.4 | 8.5 | 9.9 |
| Pt | 12.3 | 11.6 | 10.3 | 10.1 | 10.3 |
| R factor (%) | 14.3 | 12.2 | 10.5 | 10.2 | 10.2 |
| **Precision estimation** | | | | | |
| Standard deviation of Gaussian kernel (Å) | 0.48 | 0.50 | 0.51 | 0.51 | 0.52 |
| Accuracy of atom identification (%) | 97.3 | 97.7 | 98.4 | 97.3 | 96.5 |
| RMSD (pm) | 27.7 | 27.7 | 20.7 | 27.6 | 27.3 |